\documentclass{aa}  
\usepackage{graphicx}
\usepackage{txfonts}
\begin{document}
   \title{The Large Quasar Reference Frame (LQRF) - an optical representation of the ICRS}
   

   \author{A.H. Andrei
          \inst{1,2,3}
          \and
          J. Souchay\inst{3}
          \and
          N. Zacharias\inst{4}
          \and
          R.L. Smart\inst{5}
          \and
          R. Vieira Martins\inst{1,2}
          \and
          D.N. da Silva Neto\inst{2,6}
          \and
          J.I.B. Camargo\inst{2}
          \and
          M. Assafin\inst{2}
          \and
          C. Barache\inst{3}
          \and
          S. Bouquillon\inst{3}
          \and
          J.L. Penna\inst{1}
          \and
          F. Taris\inst{3}
          }

   \offprints{A.H. Andrei}

   \institute{Observat\'orio Nacional/MCT,
              R. Gal. Jos\'e Cristino 77, CEP20921-400, RJ, Brasil\\
              \email{oat1@on.br}
         \and
   Observat\'orio do Valongo/UFRJ,
              Ladeira Pedro Ant\^onio 43, CEP20080-090, RJ, Brasil
         \and
   Observatoire de Paris/SYRTE, 
              61 Avenue de l'Observatoire, F-75014 Paris
         \and
   US Naval Observatory, 
              3450 Massachusetts Av. NW, Washington, DC 20392
         \and
   INAF/Osservat\'orio Astronomico di Torino,
              Strada Osservat\'orio 20, 10025 Pino Torinese
         \and
   Centro Universit\'ario Estadual da Zona Oeste,
              Av. Manuel Caldeira de Alvarenga 1203, CEP23070-200, RJ, Brasil
             }

   \date{Received September 15, 1996, accepted March 16, 1997}

 
    \abstract
          {The large number and all-sky distribution of quasars from different  surveys, along with their presence in large, deep astrometric catalogs,  enables the building of an optical materialization of the ICRS following its defining principles. Namely, kinematically non-rotating with respect to the ensemble of distant  extragalactic objects; aligned with the mean equator and dynamical equinox  of J2000; and realized by a list of adopted coordinates of extragalatic  sources.}
          {The LQRF was built with the care of avoiding incorrect matches of its constituents quasars,  homogenizing the astrometry from the different catalogs and lists in which the constituent quasars are identified, and attaining global alignment with the ICRF, as well as fully consistent source positions}
          {Starting from the updated and presumably complete LQAC list of QSOs, the initial optical positions of those quasars are found in the USNO B1.0 and GSC2.3 catalogs, and from the SDSS DR5. The initial positions are next placed onto UCAC2-based 
       reference frames, following by an alignment with the ICRF, to which were added 
       the most precise sources from the VLBA and VLA calibrator lists - when reliable optical counterparts exist. Finally, the LQRF axes are inspected through spherical harmonics, to define right ascension, declination and magnitude terms.}
          {The LQRF contains 100,165 quasars, well represented across the whole sky, with 10 arcmin being the average distance between adjacent elements. The global alignment with the ICRF is 1.5 mas, and the individual position accuracies are represented by a Poisson distribution peaking at 139 mas in right ascension and 130 mas in declination. }
          {The LQRF contains J2000 referred equatorial coordinates, and is complemented by redshift and photometry information from the LQAC. It is designed to be an astrometric frame, but it is also the basis for the GAIA mission initial quasars' list, and can be used as a test bench for quasars' space distribution and luminosity function studies.}


   \titlerunning{The LQRF - an optical representation of the ICRS}
   \maketitle
%

\section{Introduction}

The establishment of the ICRS as the celestial reference system 
(Arias et al., 1995, Feissel \& Mignard, 1998)
answers not only the desiderata in terms of the choice of the physical
model and coherence, but also of practicality when confronted with
the earlier dynamical model. This is especially true since the system is materialized
by the direction of quasi-inertial grid points, which are given by 
distant, extragalactic sources, instead of nearby, moving stars. 
However, for a long time after the establishment of the ICRS, the
faintness of the extragalactic sources prevented a
precise determination of optical positions for a large number of them.
The VLBI radio technique answered this need, by providing precise
positions. But again, such precise positions could only be derived for a
limited number of strong and stable sources, and the ones with richer
observational history have come to define the ICRF (Ma et al. 1998). Yet even this small
number of sources enabled the VLBI to make important contributions in defining
both geophysics and radio astronomy standards. 

In contrast, concerning extragalactic sources on the optical domain, 
poor astrometry and small
number of known sources made with which the standard representation of the
ICRF were assigned to the stellar catalog Hipparcos (Perryman et al., 1997). Currently
several extensions of the HCRF (Hipparcos Catalog Reference Frame),
densify this representation, reaching fainter magnitudes.

The situation concerning the quantity of optically recognized bright 
quasars as well as concerning their sky distribution has witnessed significant  
changes in recent 
years (V\'eron-Cetty \& V\'eron, 2006, Souchay et al., 2008). 
At the same time, the science's demand for accurate astrometry of these
objects has also increased. Examples include micro and macro lensing, binaries,
and space density counts, as well as the requirements of space astronomy missions
(Andrei et al., 2008) and the very improvement of the stellar catalogs,
that requires for a dense mesh of fiducial points (Fienga \& Andrei, 2002, 2004). 

Therefore, the conditions are put together to combine the large number of quasars
optically recognized, with the large, precise optical catalogs in order to
produce an optical materialization of the ICRS in terms of its first principles,
namely kinematically non-rotating with respect to the ensemble of distant
extragalactic objects, aligned with the mean equator and dynamical equinox
of J2000, and realized by a list of adopted coordinates of extragalatic
sources. This article presents one such realization, hereafter termed 
Large Quasar Reference Frame (LQRF).

In the following section the input data is presented and reviewed, whereas in its
final subsection a schematic of the data stream is presented. The data
stream is detailed in Sects. 3, 4 and 5, where the analytical
expressions used to derive the catalog's intermediary steps are shown. 
In Sect. 3 a local astrometric solution is employed to place all
quasar input catalogs onto the UCAC reference frame. In Sect. 4
the positions derived in this way are aligned with the ICRF frame. In Sect. 5
local inhomogeneities are tackled by spherical harmonics fitting
and local average redressing. The LQRF catalog is presented in Sect. 6,
and a summary and follow-up perspectives are presented in the last section.


\section{Input data and data flow}

The LQRF is built from gathering, comparing, and adjusting the positions of 
quasars and their stellar neighborhood, extracted from several catalogs 
and lists. A brief summary of their content is presented in the following 
subsections. The role of each one is also mentioned, for the sake of 
introducing the general flow of the data treatment. The final subsection 
outlines the data flow, leaving the details for the following main sections.

\subsection{LQAC}

The Large Quasar Astrometric catalog (Souchay et al., 2008), hereafter LQAC, 
compiles the detections of 113,653 quasars from the original entries of the eight
largest lists, with the surplus extracted from the latest edition of the V\'eron \&
V\'eron catalog (V\'eron-Cetty \& V\'eron, 2006) after an analysis on the
reliability of the input contributors. The LQAC contains extensive information on the 
physical properties of each object. In particular,  it contains, whenever available,
the redshift, the magnitudes of the objects at 9 different optical bands ($u$, $b$, 
$v$, $g$, $r$, $i$, $z$, J, and K), and the radio fluxes at 5 different frequencies ($1.4 Ghz$, 
$2.3 Ghz$, $5.0 Ghz$, $8.4 Ghz$, and $24 Ghz$). The position of each
object is given by the most precise equatorial coordinate directly available from the input lists. 
Therefore, the precision is quite variable, ranging from the sub-milli-arcsec of VLBI
determinations to the level of a tenth of arcsec of the optical determinations,
up to reach nearly one arcsec of single dish determinations, On the other
hand, there are no double entries, either from misrecognition or from double radio
spots. Though the positions unevenness prevent them from characterizing an astrometric
reference frame, the LQAC positions and magnitudes, are adopted as starting point 
for finding the objects in the dense optical catalogs presented next. Figure 1
presents the LQAC sky distribution and sky density.

   \begin{figure}
   \centering
    \includegraphics[width=10cm]{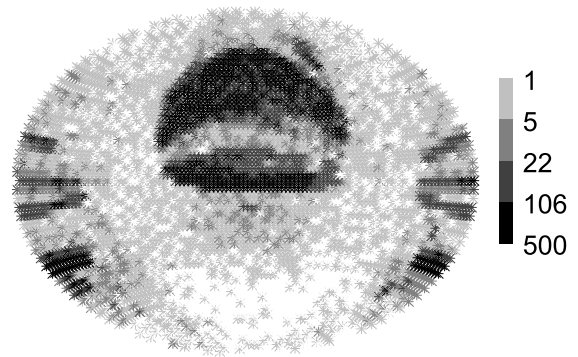}
   \caption{Sky distribution (equatorial coordinates, N up, E right) of the quasars 
   found in the LQAC catalog. The highest
density regions indicate the SDSS DR5 contribution.The scale represents the number
of sources in regions of 9 sq$^{\circ}$ shown on the plots.}
              \label{Fig1}%
    \end{figure}

\subsection{USNO B1.0}

The U.S. Naval Observatory B1.0 catalog (Monet et al., 2003), hereafter B1.0, 
is the latest one of the
series of all-sky ultra dense catalogs issued by the USNO.
It brings two epochs positions and three colors magnitudes (Johnston B, R and I) 
for 1,042,618,261 objects, up to about V=20. It results from scans of 7,435 Schmidt 
plates taken for six sky surveys and calibrated using Tycho-2 stars (Hog et al., 2000). 
The nominal
astrometric accuracy is 200 mas, but the individual objects position precision
has a median value of 120 mas along each direction of measurement. Noticing that the 
quasars forming the LQRF seat on the B1.0 fainter end, where astrometric precision 
and  accuracy degrades, the LQAC objects were searched for in the B1.0, and a
neighborhood of stars, i.e, their position, proper motion, and magnitude, was also 
collected around the found objects. Figure 2 presents the sky distribution and sky density of the 83,980 quasars recognized in the USNO B1.0

   \begin{figure}
   \centering
    \includegraphics[width=10cm]{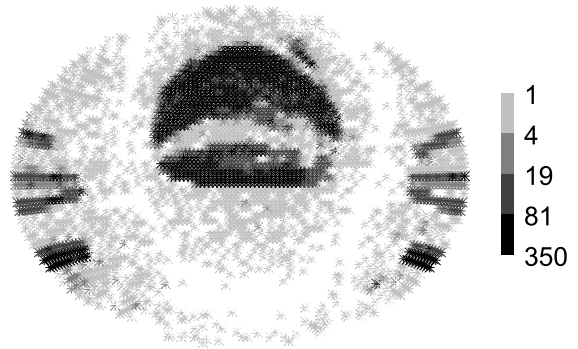}
   \caption{Sky distribution (equatorial coordinates, N up, E right) of the quasars found in the USNO B1.0 catalogs. 
 Again, the highest
density regions indicate the SDSS DR5 contribution.The scale represents the number
of sources in regions of 9 sq$^{\circ}$ shown on the plots.
Notice that the scale
is slightly different from that of Fig. 1 relative to the LQAC.}
              \label{Fig2}%
    \end{figure}

\subsection{GSC2.3}

The Second Generation Guide Star catalog latest version (Lasker et al., 2008),
hereafter GSC2.3, is an all-sky catalog derived from the Digitized Sky Survey 
(Taff et al., 1990) that
the Space Telescope Science Institute and the Osservat\'orio di Torino have created
from the Palomar and UK Schimdt survey plates. It contains position, proper
motion, and magnitude (Johnston B, R, and I) for 945,592,853 objects,
and is expected to be complete to R=20. The astrometric total error is quoted as 
smaller than 300 mas, while the relative astrometric error is better than 200 mas.
In fact the astrometric residuals against external reference catalogs only
become larger than 150 mas above R=18 at the fainter end. The astrometric
calibration was done using Tycho-2 stars. As seen, there are overwhelming similarities
between the GSC2.3 and the B1.0 raw data, which however were measured and treated by
distinct methods, enabling a most natural combination of the independent results
issued from each of them. So, as before, the LQAC objects were searched for in the 
GSC2.3, and a neighborhood of stars (position, proper motion, and magnitude) was 
collected around the found objects. Figure 3 presents the sky distribution and sky 
density of the 93,943 quasars recognized in the GSC2.3.

   \begin{figure}
   \centering
    \includegraphics[width=10cm]{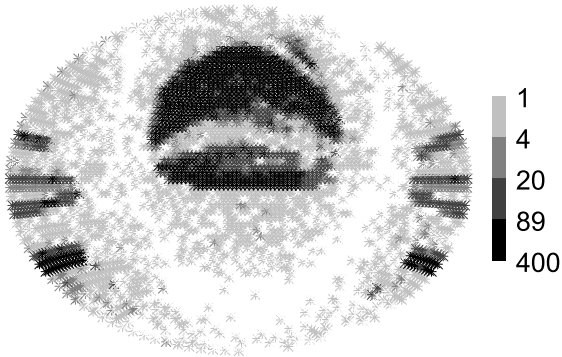}
   \caption{Sky distribution (equatorial coordinates, N up, E right) of the quasars 
   found in the GSC2.3 catalog. 
As before, the highest
density regions mark the SDSS DR5 contribution.The scale represents the number
of sources in regions of 9 deg$^2$ shown on the plots. Notice that the scale
is slightly different from that of Fig. 2 relative to the B1.0 catalog.}
              \label{Fig3}%
    \end{figure}

\subsection{SDSS DR5}

The Sloan Digitized Sky Survey (SDSS) data release 5 (Adelman-McCarthy et al. 2007),
hereafter DR5, represents the completion of the SDSS-I project. It covers 8,000 sq$^{\circ}$ ,
on the north Galactic cap (b $<$ 30$^\circ$) and on three stripes towards the 
south Galactic cap, reaching up to i=20.2, to a total of 215 million unique objects.
The astrometry is preferably referred to the UCAC catalog, or else to the Tycho-2 
catalog. The RMS residuals per coordinate are 45 mas and 75 mas for reductions
against the UCAC (Zacharias et al., 2000) and the Tycho-2 (Hoeg et al., 1997), respectively, 
with additional 30 mas systematic
errors in both cases (Pier et al., 2003). The internal astrometric precision is mainly 
limited by seeing effects, and it is quoted around 35 mas in each coordinate. A
point to notice is that quasars are signaled in the DR5, to a total of 74,869
entries with quasi-stellar object spectral class. As seen from Fig. 1, the large number of
DR5 quasars represents an important feature of the LQAC input.

\subsection{UCAC2}

The second USNO CCD Astrograph catalog (Zacharias et al., 2004), hereafter UCAC2, 
is the latest release of the ongoing UCAC project, designed 
to observe the entire sky for R magnitudes between about 7.5 and 16.  The 
observed positional errors are about 20 mas for the stars in the 10 
to 14 magnitude range, and about 70 mas at the limiting magnitude 
of R=16. The UCAC2 is a high density, highly accurate, astrometric catalog 
(positions and proper motions) of 48,330,571 stars covering the sky from -90 to 
+40$^{\circ}$   in declination and going up to +52$^{\circ}$   in some areas.  The northern 
limit is a function of right ascension. The astrometry provided in the UCAC2 is on 
the Hipparcos system. Positions are given at the standard epoch of Julian date 2000.0.
The UCAC2 supersedes the Tycho-2 catalog, having a stellar density that is over 20 
times higher. It is used here as the choice catalog to place the quasar positions
collected from the catalogs described in the above subsections onto the
Hipparcos system. Nevertheless, since the UCAC2 coverage does not attain the
entire sky, two additional stellar frames are complementary used and are presented below..

\subsection{UCACN}

What is hereafter termed UCACN is a cutout of the preliminary UCAC positions
around the LQAC quasars in the northern part of the sky that is not reached by the UCAC2
coverage (Zacharias, 2007). Although, as discussed further on, the astrometric 
accuracy of the
UCACN positions and its stellar density granted no degradation on the results
obtained using it, it will be dropped in future revisions of this work as
soon as of the forthcoming release of the all-sky catalog UCAC3. The nominal
astrometric accuracy of the UCACN stars is similar to that of the UCAC2 ones,
but no proper motion data were released for these stars. Throughout this paper when 
referring to the UCAC2 and UCACN ensemble the designation UCAC is used.

\subsection{2MASS}

The  Two Micron All-Sky Survey point source catalog (Cutri et al.,2003), hereafter 
2MASS, derives from uniformly scanning the entire sky in three near-infrared bands
to detect and characterize point sources brighter than about 1 mJy in each band, with 
signal-to-noise ratio greater than 10, using a pixel size of 2.0", 
each point in the survey having been imaged six times. The detectors worked
to a 3$\sigma$ limiting sensitivity of J=17.1, H=16.4, and K=15.3.
The 2MASS contains the position of 470,992,970 sources, but no proper motions.
The astrometry is referred to the Tycho-2 catalog and it is accurate to 70-80 mas 
over the magnitude range of 9 $<$ K $<$ 14 mag. Comparative
studies have shown that the 2MASS positions agree with those from common UCAC2
stars to within 10 mas (Zacharias et al., 2003), and that the 2MASS is compliant 
to the ICRF in the range 100-120 mas, Taking advantage of its all-sky coverage,
very high stellar density, and consistent astrometry, the 2MASS catalog is used here as 
an independent vehicle to place the LQAC quasars recognized in the large, deep
catalogs on the Hipparcos system. 

\subsection{ICRF-Ext2}

The International Celestial Reference Frame $2^{nd}$ extension (Ma et al.,1998,
Fey et al., 2004), hereafter ICRF-Ext2,
is the present materialization of the International Celestial Reference System 
(ICRS) at radio frequencies. It represents the basic frame with respect 
to which the position of any object in the celestial sphere should
be measured. In its primary contents, the ICRF consists of 212 sources, called 
"defining", whose positions are independent of the classical planes
(equator, ecliptic) and reference points (equinox), but consistent with the previous 
realizations of the Celestial System as the FK5. With the help of
a considerable amount  of VLBI observations, the individual positions of the 
sources were found to be accurate to within roughly 0.25 mas,
while the stability of the reference axes attains a remarkable 20 $\mu$as 
(micro-arcsec) accuracy. 
In the ICRF initial version 608 sources complemented the defining 212 sources.
The entrance of the ICRF-Ext1 and of the ICRF-Ext2 brought 109 new 
sources in the catalog, thus leading to a 
total number of 717 radio-sources. Notice that in the ICRF-Ext.2 catalog, a 
very small sample of objects are not quasars: 10 of them are cathegorized as 
AGN (Active Galactic Nuclei) and 10 of them are cathegorized as BL LAC (BL Lacertae). 
Although the LQAC a priori exclusively considers quasars for its compilation, these
particular objects were retained thanks to their astrometric accuracy.
The ICRF-Ext2 is used here to match the VLBI positions directly to the derived
optical positions. Such match serves to two distinct steps: (a) to redress the global 
frame orientation towards the ICRF, and (b) to redress local departures of the
derived quasar catalog from the ICRF. 

\subsection{VCS6}

The Very Long Baseline Array (VLBA) Calibrator Survey (Petrov et al., 2006),
hereafter VCS6, consists of a catalog containing milli-arcsec accurate positions 
of 3,910 extragalacic radio sources, mainly quasars.  These positions have 
been derived from astrometric analysis of dual-frequency 2.3 and 8.4 GHz VLBA 
observations, on the framework of the International Celestial Reference System (ICRS) 
as realized by the ICRF. Thus, to the level of precision and accuracy of the
optical positions determined here, the VCS6 sources represent the ICRF. In order
to strictly validate such representation, 524 VCS6 sources with formal inflated 
errors larger than 10 mas were trimmed off. The VCS6 sources not belonging to the
ICRF-Ext2 were added to it, to obtain an enlarged radio frame, which is used for
the two steps mentioned in the ICRF-Ext2 subsection, that is, the global and local redressing
of the optical quasar positions towards the ICRF.

\subsection{VLAC}

The Very Large Array (VLA) Calibrator list (Claussen, 2006), hereafter VLAC, consists
of 5,523 radio flux, structure code, position, and associated error entries for 1,860 
extragalactic sources. It is foremostly indicated as a reference tool for observations
at the VLA, but the radio positions are accurate and compliant with the ICRF. Then,
likewise in the previous subsection, the VLAC positions were added to the ICRF-Ext2
core data set, after trimming off 297 of them with errors larger than 10 mas. The combination
of the ICRF-Ext2, VCS6, and VLAC will be here termed as the enlarged radio frame (ERF)
and it represents the ICRF to within an uncertainty of 10 mas. The order of priority 
for the radio lists to compound the enlarged radio frame was, ICRF-Ext2 (718 sources),
VCS6 (2,684 sources),
and finally VLAC (59 sources). The 3,461 sources now represent the ICRF on a
better distributed all-sky basis (Fig. 4), and those for which the optical
counterpart position is determined will actually define the best adherence 
of the LQRF to the ICRF.

   \begin{figure}
   \centering
   \includegraphics[width=10cm]{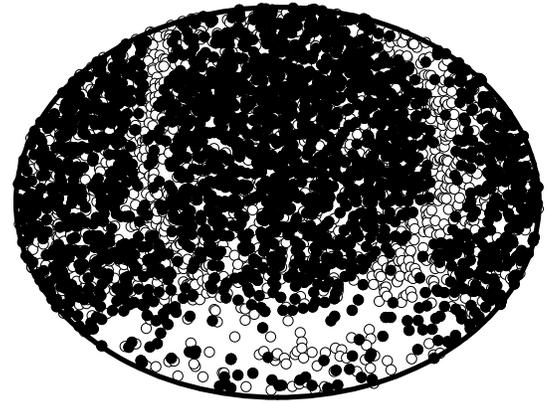}
   \caption{Sky distribution (equatorial coordinates, N up, E right) of the enlarged radio frame quasars. The filled symbols mark the sources for which an optical counterpart is found in
   at least one of the deep, dense source catalogs (B1.0, GSC2.3, DR5).
   The enlarged radio frame is formed, in order of priority of choice, by the ICRF-Ext2 (718 sources), the VCS6 (2,684 sources), and the VLAC (59 sources).
}
              \label{Fig4}%
    \end{figure}

\subsection{Data flow}

The various origins of the sources compiled in the LQAC imposes first of all
to refer their positions to a single frame in the ICRS system. After that, the
quasar frame so formed itself is aligned with the ICRF. Finally, the position of
each source is calculated by weighted average. This data flow can be sketched
as follows.

   \begin{enumerate}
   
      \item The LQAC entries are all admitted.
      \item The LQAC quasars are searched for, by a main criterion of position
coincidence, in the B1.0, GSC2.3, and SDSS DR5 high density catalogs. Each 
individual match is kept and treated separately.
      \item Around each of the individual quasars thus matched a stellar neighborhood is extracted
from the pertinent high density catalogs.
      \item The stellar neighborhood contents from the high density catalogs are
searched for in the stellar catalogs UCAC2, UCACN, and 2MASS, as before by a main 
criterion of position coincidence.
      \item Within each neighborhood, the stars common to the high density catalog
and to the stellar catalog have their positions projected onto the tangential plane
centered at the quasar. A local transformation function from the high density
catalog to the stellar catalog is calculated, mimicking the complete plate
solution polynomials.
      \item At this point, for each quasar, typically, a number of positions are
determined (e.g., from more than one high density catalog, and local solutions 
from the two stellar catalogs). Each of these families of positions will still be 
handled separately.
      \item Within each family, the quasars for which there is a position in the 
enlarged radio frame (that is, the combination of the ICRF, VCS6, and VLAC catalogs) 
are found, since they are so flagged in the LQAC, what as before follows mainly a criterion
of positional coincidence. 
      \item The subsets so defined are used to calculate the global rotation and
zero point corrections to obtain coincidence with the ICRF origin and axes. The
global rotation and zero point corrections are next applied to the positions of
all quasars of each given family.
      \item Likewise the previous step, the optical and radio positions in
subsets are combined using orthogonal functions, in right ascension, declination,
and magnitude, in order to correct systematic local departures from the
ICRF axes directions. And likewise the significant orthogonal functions are
applied to the positions of all quasars of each given family.
      \item Again the optical and radio positions in
subsets are combined in small regions around each quasar to derive its correction
for localized inhomogeneities.
      \item For each position of each source in a given family, a total error is
assigned that combines the formal error from the high density catalog where it appears,
the error from the particular local correction, the error from the particular
global rotation and bias, and the error from both the significant orthogonal function 
adjustments and the inhomogeneity corrections. 
      \item From the squared inverse of the total error, a weight is assigned to
each determined position of a given quasar. The final position of each quasar is 
given as the weighted average of the determined positions. The error in the position
is assigned according to an individualized function of the optical to radio
disagreement.

   \end{enumerate}

The comparisons of the final positions against the existing corresponding positions
in the enlarged frame is used to obtain the global properties of the LQRF.


\section{The local astrometric solution}

In this section the points 1 to 6 of Sect. 2.11 (data flow) are
derived.

Starting from the LQAC input list of quasars, the first step to get
their astrometric positions, in order to form a whole consistent reference frame,
is to have them matched to the position of objects in the two large,
deep all sky catalogs, the USNO B1.0 and the GSC2.3, as well as in the SDSS DR5. 
The basic matching criteria is of position agreement within 1 arcsec. 
This is equal to the poorest astrometric accuracy in the LQAC 
contributing surveys. It is also a safe threshold given the typical
seeing and astrographic plate resolution limits. Additionally a
magnitude limit was imposed preventing the selection of objects
brighter than R=7, nevertheless this threshold was not reached by any source. 
No proper motion limit was enforced because both the USNO B1.0 and the GSC2.3
warn against possible offsets of the proper motions zero
points. A post check was made by determining local zero points of proper
motion, given by the average motion. In comparison to the local zeros
of proper motion, just 0.7\% of the GSC2.3 selected objects show significant
motion, while for the B1.0 selected objects just 0.1\% of them
show significant motion. To verify what would be the possibility of a
chance hit, the input coordinates were varied by random values between
1 and 5 arcmin. In this way the same region of the true objects
was being swept. Keeping the search radius to 1 arcsec, in only 
0.2\% of the cases was verified a hit on the false coordinates.
Moreover, in only 0.1\% of cases was a second object found significantly
close to the adopted match. These satisfying proportions change very little
when the search radius is enlarged to 2 arcsec. While this is
an added proof of the overwhelming correctness of the matches, it at
the same time indicates that there is little advantage in adopting a larger 
radius, and certainly no gain to the confidence level attached to the
cross-identifications.

In all, 83,980 quasars are recognized in the B1.0, and 93,943 quasars
are recognized in the GSC2.3. The sky distribution follows closely that
shown in Fig. 1.
From the figure it is clear the large contribution of the SDSS DR5.
For the later, 74,825 quasars are retrieved using the SDSS query server.
The positions from the three catalogs were added independently to the input list. 
The final input data contains 100,165 quasars. Just small fractions of
the total appear in only one of the catalogs (1,405 only in USNO B1.0,
4,425 only in GSC2.3, and 3,276 only in SDSS DR5), while 61,532 quasars
appear in all three catalogs.

These three catalogs that bring the input positions for the recognized
quasars are originally placed on the ICRF J2000 reference frame
by different astrometric pipelines based on the Tycho-2 catalog.
Their global orientation and equatorial bias relative to the 
ICRF J2000 frame are shown in Table 1. Local deviations are also
apparent in Figs. 5-7.

\begin{table}
\caption{Initial optical to radio relationships$^{\mathrm{a}}$.}

  \label{tab:table1}
\begin{tabular}{l r r r}
\hline \\
 Tracer & B1.0 to RadPos & GSC2.3 to RadPos & DR5 to RadPos \\
\\
\hline \\
N & 1962 & 2028 & 300 \\
$\overline{\Delta\alpha cos\delta}$ & -35.2 $\pm$ 4.7 & +32.9 $\pm$ 5.5 & -~4.5 $\pm$ 3.2 \\
$\overline{\Delta\delta}$ & +123.5 $\pm$ 4.9 & +~39.5 $\pm$ 4.7 & +~13.9 $\pm$ 3.2 \\
A1 & +4.7 $\pm$ 6.2 & -6.6 $\pm$ 6.7 & -3.5 $\pm$ 4.7 \\
A2 & +5.9 $\pm$ 5.8 & +13.1 $\pm$ 6.3 & -19.4 $\pm$ 5.1 \\
A3 & +42.1 $\pm$ 5.7 & -29.0 $\pm$ 6.1 & +6.6 $\pm$ 4.1 \\
A4N & +123.0 $\pm$ 5.3 & 20.9 $\pm$ 5.7 & 1.6 $\pm$ 4.5 \\
A4S & +128.0 $\pm$ 11.9 & 134.1 $\pm$ 12.5 & ~ \\
\\
\hline \\
\end{tabular}
\begin{list}{}{}
\item[$^{\mathrm{a}}$] For the catalogs USNO B1.0, GSC2.3, and SDSS DR5 relatively to
the collected sample of sources with precise radio interferometric positions 
(Extended Radio Frame - RadPos, either from ICRF-Ext2, VCS6 or VLAC), are presented: 
the average offsets 
in right ascension and declination, the orientation angles (A1, A2, A3) 
relative to the standard thriedron of equatorial coordinates, and
the equatorial biases (A4N and A4S, north and south of -20$^{\circ}$). 
All values are in milli-arcsec (mas) 
and the number of sources is given in the first line.
\end{list}

\end{table}

   \begin{figure*}
   \centering
   \includegraphics[width=18cm]{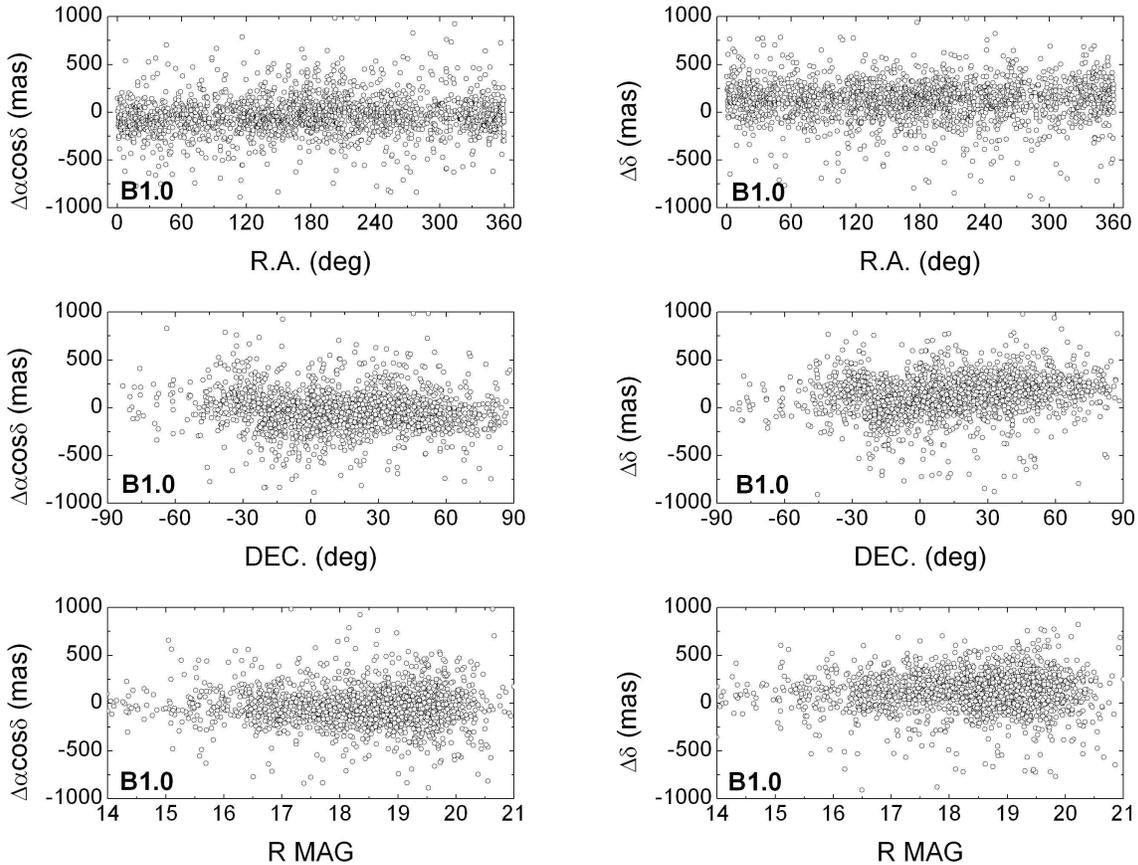}
   \caption{USNO B1.0 right ascension and declination offsets distributions
relative to the equatorial coordinates axes and the R magnitude. The
offsets are given in the sense of catalog minus the radio interferometric 
positions from the ICRF-Ext2, or the VCS6, or the VLAC.}
              \label{Fig5}%
    \end{figure*}

   \begin{figure*}
   \centering
   \includegraphics[width=18cm]{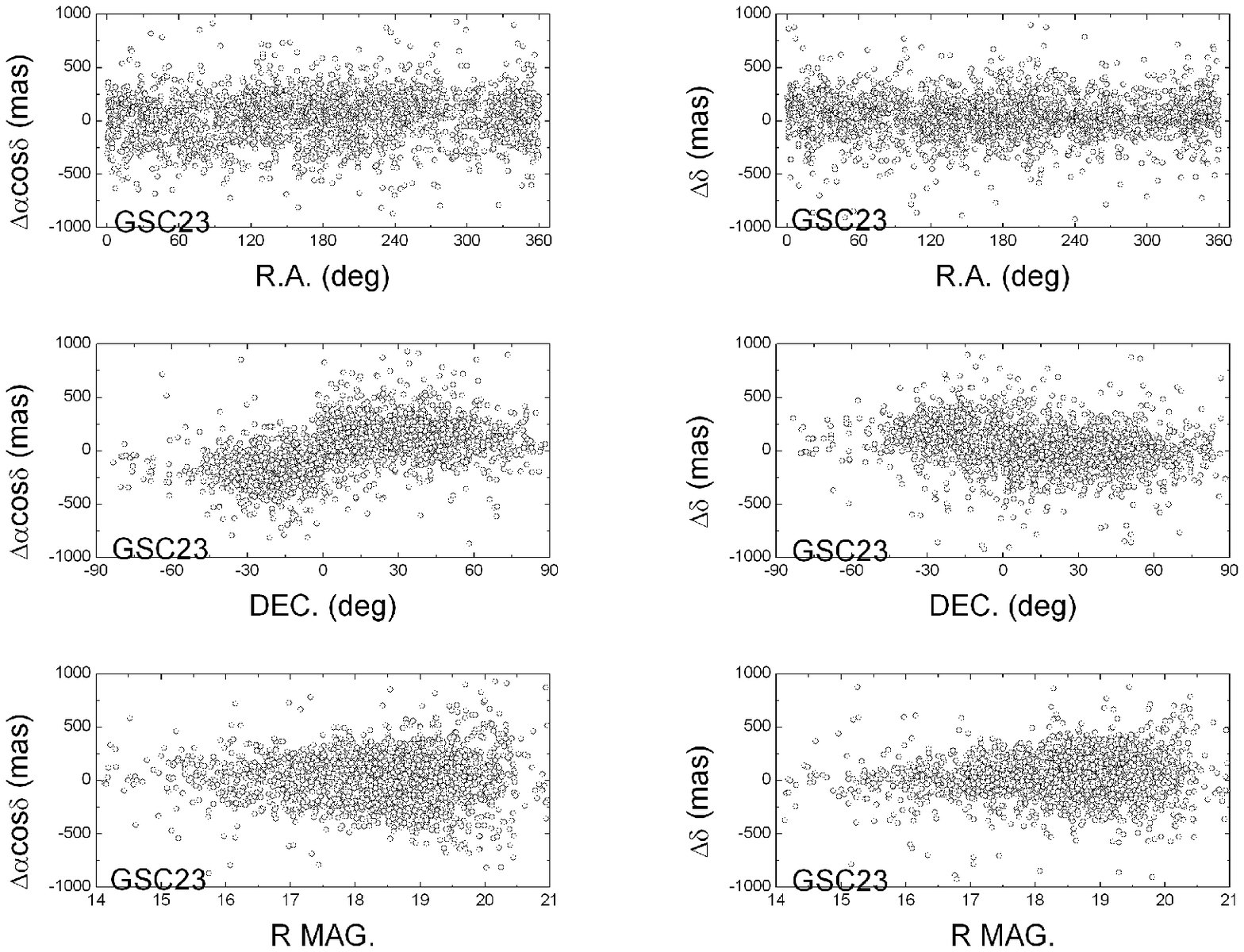}
   \caption{GSC2.3 right ascension and declination offsets distributions
relative to the equatorial coordinates axes and the R magnitude. The
offsets are given in the sense of catalog minus the radio interferometric 
positions from the ICRF-Ext2, or the VCS6, or the VLAC.}
              \label{Fig6}%
    \end{figure*}

   \begin{figure*}
   \centering
   \includegraphics[width=18cm]{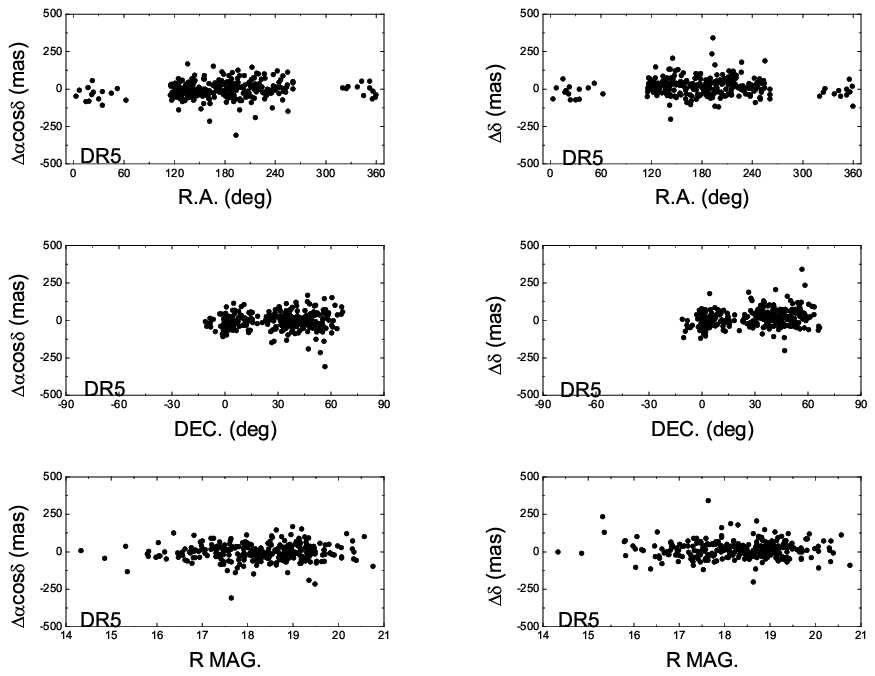}
   \caption{SDSS DR5 right ascension and declination offsets distributions
relative to the equatorial coordinates axes and the R magnitude. The
offsets are given in the sense of catalog minus the radio interferometric 
positions from the ICRF-Ext2, or the VCS6, or the VLAC.}
              \label{Fig7}%
    \end{figure*}

In order to homogenize all the original different positions, a local astrometric 
reduction was performed based on the catalogs UCAC and 2MASS. The 
local correction follows the standard tangent plane astrometric
reduction, taking the quasar original position as the central point of the 
rigorous gnomic projection (Assafin, Vieira Martins \& Andrei, 1997). 
Through it the catalogs coordinates are adjusted to the
reference frames represented by the stellar catalogs. Three assumptions
are implicit to the method. That the stellar catalogs deliver an
improved, and similar, representation of the ICRF, either relatively
to the two Schmidt plate surveys, as well as relatively to the photometric 
one. That in a very restricted neighborhood the deviation between the deep 
surveys
and the stellar catalogs, whichever one it is originated from, can be modeled
by simple relationships. And that the removal of such deviations can
significantly improve the accuracy of the quasar positions found in
the large surveys. Prior investigations support these assumptions
(da Silva Neto et al., 2005).

\subsection{The polynomial degree of the solution}

On the functional side, the initial question to be tackled is on the
polynomial degree of the local astrometric correction representation. 
The UCAC is the
least dense of the involved catalogs, and it presents a typical 
density of 10 stars per 5 $\times$ 5 arcmin. Translated to the
Schmidt plates scale this means that a first degree solution with 10
reference stars can be applied over a physical area as small as 25mm$^2$.
According to the adopted a priori assumptions, and given the sky density
of the stellar catalogs, the simplest models seem to be well suited.
To verify such procedure, a trial local reduction was performed for
29 quasars randomly distributed across the celestial sphere, at every
one hour of right ascension, off the galaxy plane. Around
each of them, a field of UCAC2 stars was selected in boxes of side 10
arcmin, which contained at least 30 stars. For these 29 fields of quasars, 
complete polynomial solutions were tested, with orders varying from 0 to
3$^{rd}$, and a straight averages solution. The quasar positions were taken
from the USNO B1.0 and all positions were transported to the USNO B1.0
plate epoch, using the proper motions of either catalog. Table 2 summarizes
the mean results.

  \begin{table}
  \caption{Comparison between the models for the local astrometric solution$^{\mathrm{a}}$.}
  
  \label{tab:table2}
          \begin{tabular}[h]{lcccc}
        \hline \\[-5pt]
        Solution &  
        $\overline{\Delta\alpha cos \delta}$ &
        $\sigma_{\Delta\alpha cos \delta}$ & 
        $\overline{\Delta\delta}$ &
        $\sigma_{\Delta\delta}$ \\[+5pt]
  
        \hline \\[-5pt]
  
        3$^{rd}$ degree  & $+$42 & 126 & $-$25 & 140 \\
        2$^{nd}$ degree  & $+$43 & 127 & $-$23 & 136 \\
        1$^{st}$ degree  & $+$61 & 110 & $-$34 & 120 \\
        0 degree         & $+$42 & 126 & $-$34 & 119 \\
        offset           & $+$56 & 118 & $-$45 & 130 \\
        \hline\hline \\[+5pt]
        Solution &  
	        $\overline{\Sigma _X}$ &
	        $\sigma_{\Sigma X}$ &
	        $\overline{\Sigma _Y}$ &
                $\sigma_{\Sigma Y}$ \\[+5pt]

        \hline \\[-5pt]
  
        3$^{rd}$ degree  & 137 & 63 & 127 & 64 \\
        2$^{nd}$ degree  & 134 & 66 & 125 & 64 \\
        1$^{st}$ degree  & 135 & 68 & 120 & 65 \\
        0 degree         & 133 & 68 & 117 & 64 \\
         \hline \\
        \end{tabular}
  \begin{list}{}{}
  \item[$^{\mathrm{a}}$] In the top half of the table, the average corrections and
  their standard deviations are presented. In the bottom half, the average internal 
  error in the solutions and their dispersion are presented.
  All values are in milli-arcsecs.
  The difference between the "0 degree" and the "offset" solutions is
  just that in the former residuals larger than 2.5$\sigma$ are removed
  and the solution is recalculated.
  \end{list}

  \end{table}

In the first half of Table 2 the average local corrections and the
corresponding standard deviation are shown. Though the values only very 
loosely represent the systematic offsets between the USNO B1.0 and the UCAC2,
what is important to note is the similarity between all the rows. In the second half 
of the table, the average internal errors of each solution and their
dispersion are shown. Again all rows behave quite alike. Therefore, to
keep the smallest the region where the local solution is derived, and to
maximize the ratio of the number of equations to the number of
parameters to find, the complete first degree polynomial was adopted.
This solution also produces the smallest set of standard deviations.
The zero degree solutions fared just as well as the others, so there would
be no particular advantage on taking an even smaller region, while
the first degree polynomial allows for some modeling, which is useful face to
the large numbers of sources that the local solution is going to be applied to.
Two formulations of the first degree polynomial are concurrently used,
the complete one (six independent parameters) - hereafter the six parameter model, 
and the so-called four constants one (in which the X and Y axes are 
constrained to have the same scale) - hereafter the four parameter model. 
Furthermore, notice that with the six parameter model the
error in the central, pseudo-tangential point, which is always the
position of the quasar, was also found to be the smallest.

\subsection{The stellar catalogs and the size of the local segion}

As discussed above, the UCAC2 offers the desiderata required for the local
astrometric solution: high stellar density, accurate global representation
of the HCRF and hence of the ICRF, range of magnitudes up to 16, and small
zonal and local frame biases. The stellar density supports using a region
as small as 10 arcmin across, which is always within the overlap zone
of the Schmidt plate surveys contributing to the B1.0 and the GSC2.3. 

However, the UCAC2 has incomplete coverage in the northern celestial hemisphere. 
Two remedies are used to face such hindrance, the use of
preliminary positions of northern UCAC stars (here called UCACN), and the use
of the 2MASS catalog. In the former case, the UCACN positions are
by construction on the same system of the UCAC2, and at the same precision.

The 2MASS Point Source catalog positions are compliant
with the ICRS via the Tycho-2 catalog and expected to be accurate to 70-80 mas 
over the magnitude range of 9 $<$ Ks $<$ 14 mag. At the limit Ks = 16, the 
accuracy is not better than 200 mas. As the 2MASS does
not provide proper motions, these values are good for the mean epoch of observation,
1999.

The pitfalls due to the magnitude gap between the 2MASS objects and the
brighter Tycho-2 reference stars degrade the final astrometric accuracy. The 
internal precision of the 2MASS individual positions is better, at the
level of 40-50 mas in the 9 $<$ Ks $<$ 14 mag range. The comparison between
5.2 million stars common to the UCAC2 shows standard deviations of 70-80 mas
in both coordinates, for the 9 $<$ Ks $<$ 14 mag range. For the common stars
around the quasars extracted to define the local solution, the difference average 
and standard deviation are $\overline{\Delta\alpha cos \delta}$ = $+$3 mas,
$\sigma_{\Delta\alpha cos \delta}$ = 99 mas, and $\overline{\Delta\delta}$ = $+$11 mas,
$\sigma_{\Delta\delta}$ = 95 mas, all in the sense 2MASS.minus UCAC2. Furthermore,
from the limited sample of 700 quasars from the enlarged radio frame found in the
2MASS catalog, the analogous statistics are 
$\overline{\Delta\alpha cos \delta}$ = $+$0 mas,
$\sigma_{\Delta\alpha cos \delta}$ = 144 mas, and $\overline{\Delta\delta}$ = $-$2 mas,
$\sigma_{\Delta\delta}$ = 138 mas.

On the positive side, for the same region, the 2MASS catalog presents about 10 times the
number of reference stars that does the UCAC catalog. In the small regions used
for the local solution, it represents a coherent local frame, from which the zonal
bias can be removed after later comparisons against the enlarged radio frame
sources. It is well aligned with the UCAC and with the ICRF, and represents
an all-sky frame. The local solution was computed independently using UCAC and
2MASS reference stars. Each of the results contribute, again independently, to the
weighted average used to compute the final LQRF quasar coordinates.

\subsection{The analytical expression for the local correction astrometric solution}

On establishing the analytical form of the local astrometric solution, a final point 
to be considered concerns the usefulness of a magnitude dependent term, 
a feature often present in Schmidt plates based surveys (da Silva Neto et al., 2000).
In the present case, the R magnitude represents the obvious choice, mainly
because it is common to all the involved catalogs, and closer to the 
central wavelengths of the 2MASS bands. At the same time, due to the quasars redshift
and 
to the reddening for the stars, coupled with CCD top sensitivity and wider
plate effective passband, the signal-to-noise ratio is usually higher towards the red. 
An inspection on the magnitude dependency of the catalog minus enlarged
radio frame residuals
reveals a small linear trend, larger for the $\Delta\delta$ offsets
than for the ${\Delta\alpha cos\delta}$ offsets. The same trend is
verified for both B1.0 and GSC2,3 increasing with magnitude for 
declination, and decreasing with magnitude for right ascension.  The 
amplitudes get only to 10 mas in the significant cases. In what concerns 
the DR5, the magnitude dependence is of much smaller amplitude and opposite 
tendency. The possibility of a magnitude equation is not ruled out in the 
presentation of either of the two deep catalogs. However, the causes and 
ensuing description would be out of the scope of this work. Here, an
additional obstacle would be to bridge the magnitude gap between the
reference stars and the much dimmer quasars. All accounted, the magnitude
investigation is left to be investigated by the orthogonal functions in
direct comparison with the enlarged radio frame positions.

Therefore, the enforced local solutions solution are:

\begin{eqnarray}
\xi_{SCat} - X_{QCat} &= &a\,X_{QCat} + b\,Y_{QCat} + c
\label{eq:equation1}
\end{eqnarray}
\begin{eqnarray}
\eta_{SCat} - Y_{QCat} &= &d\,Y_{QCat} + e\,X_{QCat} + f
\label{eq:equation2}
\end{eqnarray}\\

In Eqs (1) and (2), "SCat" is either the UCAC2 (or even its
preliminary northernmost part) or the 2MASS, and "QCat" designates
the B1.0, the GSC2.3 or the DR5. The relations
above are fit to the smallest set of common "SCat" and "QCat"
stellar objects in a neighborhood larger than 10 arcmin but smaller 
than a box of size of
30 arcmin, to a minimum of 6 stars. The solutions are calculated
independently for right ascension and declination. Whenever that minimum number of stars  
is touched, either at the start or after elimination of stars, the four parameter
model is used - in that case, obviously combining the two
equatorial coordinates solutions. In fact, as presented in the next 
section, the four parameter model was run for all quasars and
it delivers solutions in excellent coincidence with the six parameter
solutions. The analytical form of the four parameter model is presented
in Eqs. (3) and (4), where "SCat" and "QCAT" keep their afore-defined meaning.

\begin{eqnarray}
\xi_{SCat} - X_{QCat} &= &g\,X_{QCat} + h\,Y_{QCat} + i
\label{eq:equation3}
\end{eqnarray}
\begin{eqnarray}
\eta_{SCat} - Y_{QCat} &= &h\,Y_{QCat} - g\,X_{QCat} + j
\label{eq:equation4}
\end{eqnarray}\\

\subsection{Applying the local astrometric solution}

As discussed in the previous sections, the initial task of producing
the LQRF consists of carrying out a local astrometric reduction using the most
precise and accurate stellar catalogs representing the HCRF,
through the Tycho-2 reference frame, in order to locally redress
the quasars' position input from the high density, deep catalogs.. 

There, however, the question
of epochs comes up. The stellar catalogs have a mean epoch very
close to J2000, being J1998.79 ($\pm$ 2.78 y) for the UCAC2, and
J2003.34 ($\pm$ 0.24 y) for the UCACN, the UCAC2 preliminary
northernmost part, and J1998.87 ($\pm$ 0.08 y) for the 2MASS. 
Also, the DR5 mean epoch is J2002.96 ($\pm$ 1.45 y), raising
no questions. The dilemma on whether the local reduction ought
to be performed in the epoch of the quasar input position or directly
in J2000 is actually only meaningful for the cases when the B1.0 and the GSC2.3 
are locally reduced by the UCAC2 . For them it would not be unreasonable
to speculate that a reduction made at the quasar epoch would diminish
the scatter in the stellar positions, on the supposition that the stellar
images themselves were at the quasar epoch (i.e., the epoch of the plate where the quasar is found) because of the smallness
of the used field. The B1.0 mean epoch is J1977.12 ($\pm$ 5.49 y).
On the other hand, it is verified that the B1.0
positions are given as an average of plates reductions, in such a way
to jeopardize the hypothesis of common epochs, given the range of
magnitudes and to some extent the two different types of objects.
For the GSC2.3, in contrast, the positions always refer to one single highest quality
plate, whenever possible the red one. But, again, the GSC2.3 mean epoch
is J1992.5 ($\pm$~3.72~y) and therefore there is a balance between
reducing modestly the stellar positions scatter and the errors brought
in by faulty or biased proper motions. 

In order to actually verify which way would be preferred, the local
corrections for the sources belonging to the enlarged radio frame
were evaluated both at J2000 and the quasar position epoch. A second trial was made
by considering only the sources from the enlarged radio frame 
belonging to the three input catalogs and that could be reduced by all
the stellar frames. This subsample contains 240 sources but enables
a direct comparison between all the possible ways of treatment. The
outcomes show no net advantage in working at the quasar epoch. 
For the B1.0 catalog, the number of sources for which the
local reduction converges drops; there is a marginal gain on the
number of stars entering in the reduction but the number of
actually used stars remains the same; there is no diminishing on the
standard deviations from the solution corrections, or from the
optical minus radio offsets; and the removal of the equatorial bias
is less efficient (by 98 mas). The external check actually shows a small
increase of the average detachment to the representation of the ICRF origin.
For the GSC2.3 there is no important
loss in working at the quasar epoch, but no gain either. The number
of sources for which the local reduction converges remains the same, 
and so does the number of entry and used stars. The standard deviation in
the solution does not change, and for the optical minus radio positions the
offset scatter improves only marginally (by 4.5 mas). The equatorial bias
is less efficiently removed (by 40 mas). Exactly the same results
are obtained when inspecting the subsample of common sources.
In view of these results it was decided to work only at J2000 for
the local corrections. Thus all pairs of input quasar catalogs and
stellar catalogs are treated identically.

A second point concerning the issue of epochs has already been touched
in the previous discussion. It concerns whether the local reductions
made with the UCACN, that is for the northernmost sources, and with 
the 2MASS would be significantly worse than those made with the UCAC2.
As seen, both of them have a mean epoch close to J2000, where the local
corrections are performed. For the 2MASS catalog, which is not originally
presented as an astrometric frame, the number of equations of condition 
is substantially increased for a same patch of the skies as defined 
for the UCAC2. To test the performance of the different catalogs, we use
the DR5, which was shown in Sect. 3 to have intrinsically smaller scatter 
relative to the ICRF.  As before, the enlarged radio frame and its subset
of all common sources are employed, and they provide responses much alike. 
The UCAC2 based reductions produce the best results as expected. 
The right ascension and declination standard deviation
in the offsets to the radio frame is at 47.5 mas. We remark that this is 
already a gain relative to the DR5 catalog offsets that are
at 59 mas, as produced from Table 1 values. The standard deviation in
the solution itself is at 33 mas. The UCACN based reductions
perform much alike, being actually slightly better on declination and
30 mas worse on right ascension. The optical minus radio standard
deviation is 14 mas worse than that of the UCAC2.  The 2MASS based 
reductions fare somewhat worse than the UCACN ones, being larger by 18 mas
in the solution standard deviation, and worse by 6.5 mas in the
optical minus radio standard deviations. In this 
case, it is worth to point out that the average optical to radio position
offsets are much similar
for the UCAC2 and the 2MASS (differing by just 5 mas). It is also to be
stressed that, eventhough not bringing a direct improvement in accuracy
as the UCAC2 based ones do, the UCACN and the 2MASS local reductions
enable to place the DR5 positions on the same frame with the
B1.0 and GSC2.3 reduced positions, without impairing their
intrinsic precision.

Table 3 summons up all the applied local corrections.

\begin{table*}[htb]
  \caption{Mean results from the local astrometric corrections$^{\mathrm{a}}$.}

  \label{tab:table3}
  \begin{center}
    \leavevmode
        \begin{tabular}[h]{lrrrrrrr}
      \hline \\[-5pt]
      Solution & Nq & $\overline{Ns}$ &
      $\overline{C\alpha cos \delta}$ & $\overline{C\delta}$ &
      Nrad & $\overline{\Delta\alpha cos \delta}$ & $\overline{\Delta\delta}$ 
       \\[+5pt]

      \hline \\[-5pt]

BU6 & 47822 & ~13.0 & -23.5 (106.2) & -145.3 (104.2) &
 1320 &   -43.5 (177.0)  & +12.2 (171.6) \\
BN6 & ~3023 & ~14.1 & +~4.2 (~45.6) & -230.0 (~81.7) &
 123 &   -70.5 (136.6)  & -~9.9 (121.6) \\
BT6 & 83951 & ~52.8 & +~3.3 (107.9) & -178.8 (111.2) &
 1897 &   -18.2 (172.9)  & -12.0 (167.5) \\
BU4 & 43991 & ~13.8 & -21.8 (105.1) & -147.8 (103.6) &
 1273 &   -41.9 (176.6)  & +11.0 (171.8) \\
BN4 & ~3095 & ~14.1 & +~4.7 (~45.8) & -229.1 (~80.3) &
 ~124 &   -68.8 (136.6)  & -11.4 (121.9) \\
BT4 & 84012 & ~52.7 & +~3.2 (107.7) & -178.9 (111.1) &
 1896 &   -18.6 (172.2)  & -12.9 (167.0) \\
GU6 & 76671 & ~22.7 & -58.4 (109.7) & -~~8.8 (~91.4) &
 1633 &   -27.2 (178.3)  & +33.7 (166.2) \\
GN6 & ~4131 & ~16.3 & -56.3 (~51.8) & -~~5.4 (~87.8) &
 ~150 &   +21.8 (148.8)  & -11.9 (136.5) \\
GT6 & 93976 & 119.1 & -55.9 (135.4) & -~27.3 (119.2) &
 1984 &   -~5.1 (162.3)  & +~0.4 (157.3) \\
GU4 & 77110 & ~22.6 & -58.2 (109.9) & -~~8.9 (~91.4) &
 1634 &   -26.5 (178.2)  & +33.0 (165.5) \\
GN4 & ~4356 & ~16.2 & -56.6 (~51.5) & +~~6.2 (~86.3) &
 ~154 &   +20.1 (150.0)  & -14.0 (133.3) \\
GT4 & 93978 & 119.1 & -55.8 (135.4) & -~27.4 (119.3) &
 1983 &   -~5.2 (161.8)  & +~~0.3 (157.4) \\
SU6 & 59184 & ~30.0 & +~6.5 (~22.3) & +~14.2 (~26.2) &
 ~241 &   +~5.7 (~44.3)  & +25.6 (~49.9) \\
SN6 & ~4035 & ~16.0 & -~7.6 (~24.2) & -~26.8 (~48.9) &
 ~~37 &   -15.0 (~45.7)  & -~1.8 (~37.7) \\
ST6 & 74838 & 167.6 & +~5.3 (~49.4) & +~20.8 (~55.5) &
 ~301 &   +~9.8 (~65.9)  & +30.1 (~63.1) \\
SU4 & 59125 & ~30.0 & +~6.4 (~21.8) & +~14.1 (~25.8) &
 ~240 &   +~6.2 (~43.3)  & +24.1 (~47.6) \\
SN4 & ~4036 & ~16.0 & -~7.5 (~23.9) & -~26.5 (~48.3) &
 ~~37 &   -14.7 (~45.4)  & -~0.8 (~37.0) \\
ST4 & 74838 & 167.6 & +~5.4 (~49.3) & +~19.9 (~55.2) &
 ~301 &   +~9.7 (~65.9)  & +29.0 (~63.3) \\

      \hline \\
      \end{tabular}
  \end{center}
 \begin{list}{}{}
 \item[$^{\mathrm{a}}$] Comparative results given by all types of local correction
           designed. The solutions nomenclature is formed by one letter
           designating the quasar position input catalog, followed
           by one letter designating the stellar reference frame, 
           and by the codes $6$ and $4$ informing whether the algorithm
           is of the six or four parameter model.
            $Nq$ is the number of quasars for which 
           a given local correction was possible and succeeded. 
           $\overline{Ns}$ is the average number of catalog stars 
           contributing to the solution.   $\overline{C\alpha cos \delta}$ and 
           $\overline{C\delta}$ columns bring the mean corrections on
           right ascension and declination, and the
           standard deviations (in brackets). The final two columns
           show the average optical to radio offset, preceded by the
           number of sources entering the average ($Nrad$). 
           All angular values are in milli-arcsec. See the text for
           detailed explanations on the solutions and samples.
 \end{list}
\end{table*}

As can be seen from Table 3, the different quasar input 
catalogs (codes B, G, or S) require different local corrections, while the three stellar reference 
catalogs (codes U, N, or T) deliver such corrections at similar accuracy. 
The local correction outputs (codes C) agree when calculated either by the
six parameter or by the four parameter models (codes 6 or 4).
The local corrections are generally
larger than the remaining optical minus radio offsets (codes $\Delta$), which supports
their usefulness. The optical minus radio mean offset in all cases is
improved relatively to the one obtained using original quasar catalogs positions. 
As a consequence, the mean offsets become similarly small.

At this point, the three quasar input catalogs and the three stellar reference
frames deliver concurrent, independent sets of positions. And whenever the
1$^{st}$ degree polynomial solution fails to converge, the four parameter
solution results can be adopted equivalently. There are thus up to
twelve possible local solutions calculated for each source.
For 61\% of the sources more than one solution was obtained, for 844 
sources all twelve solutions went through, and for 12 sources no solution
converged. Figures 8 to 11 display the local corrections for the most critical
cases, namely the B1.0 and the GSC2.3, as given by the 
UCAC and 2MASS frames. Comparing with the Figs. 5-6, it is already 
clear that the
local corrections bring the quasar's positions onto the ICRF. The
comparison between these figures and the initial plots (Figs. 5-7) 
show that the local astrometric solution significantly redress the input 
positions from the large catalogs. Notice that the outcomes from the UCAC
and 2MASS stellar reference frames bear much in common. Likewise, the
six parameter and the four parameter models derive coincident corrections
(thus, only those from the six parameter model are shown in the figures).
It is also interesting to note that there is no discontinuity between
the zones where either the UCAC2 or the UCACN stellar frames were used.

   \begin{figure*}
   \centering
   \includegraphics[width=18cm]{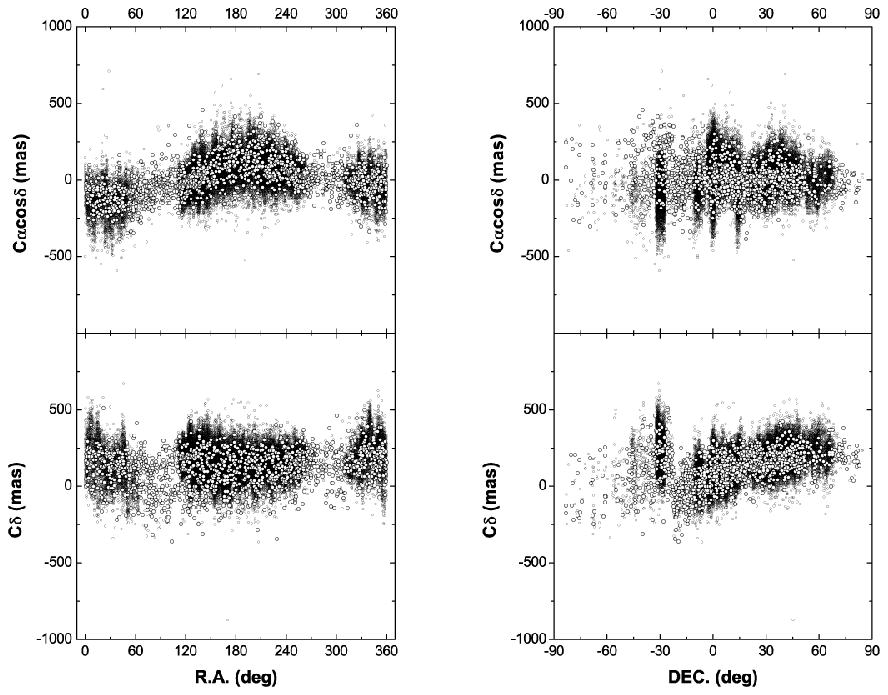}
   \caption{Equatorial coordinates distribution of the local astrometric
   corrections derived by the six constants model using the UCAC stellar
   reference frame onto the USNO B1.0 quasar positions. The clear points
   correspond to the quasars for which there are ERF radio positions.
   }
              \label{Fig8}%
    \end{figure*}

   \begin{figure*}
   \centering
   \includegraphics[width=18cm]{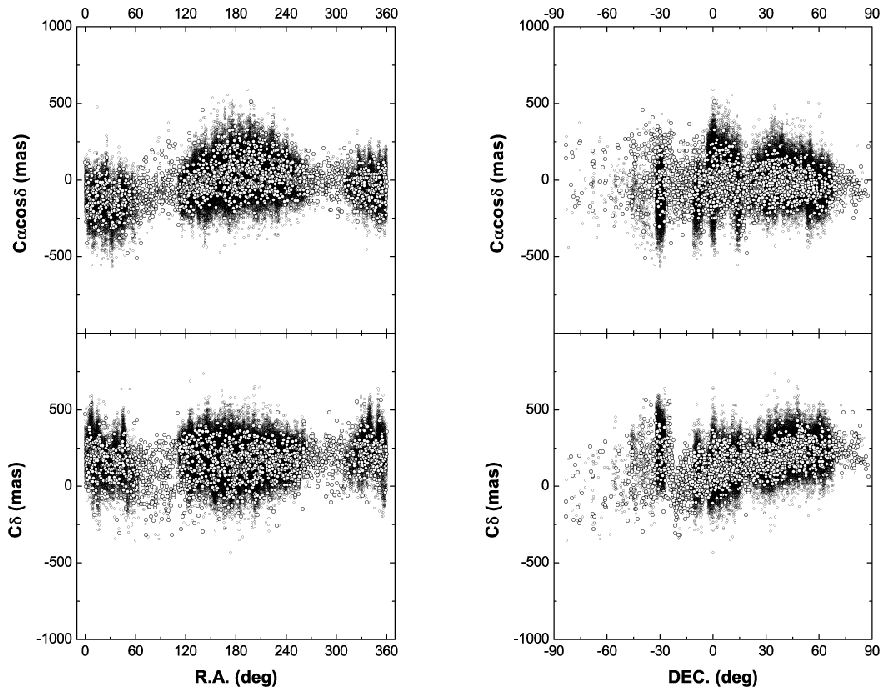}
   \caption{Equatorial coordinates distribution of the local astrometric
   corrections derived by the six constants model using the 2MASS stellar
   reference frame onto the USNO B1.0 quasar positions. The clear points
   correspond to the quasars for which there are ERF radio positions.}
              \label{Fig9}%
    \end{figure*}

   \begin{figure*}
   \centering
   \includegraphics[width=18cm]{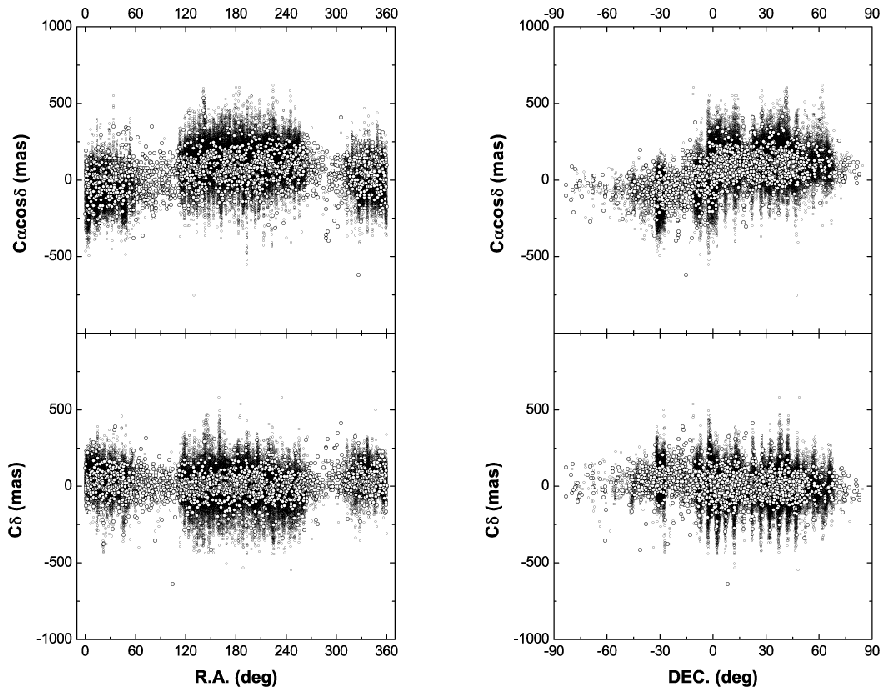}
   \caption{Equatorial coordinates distribution of the local astrometric
   corrections derived by the six parameter model using the UCAC stellar
   reference frame onto the GSC2.3 quasar positions. The clear points
   correspond to the quasars for which there are ERF radio positions.}
              \label{Fig10}%
    \end{figure*}

   \begin{figure*}
   \centering
   \includegraphics[width=18cm]{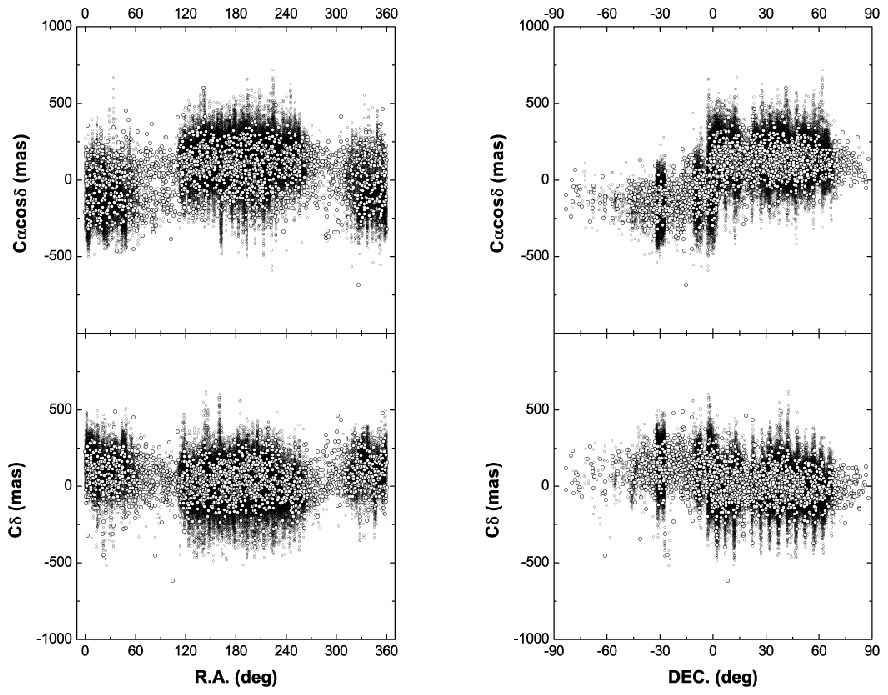}
   \caption{Equatorial coordinates distribution of the local astrometric
   corrections derived by the six parameter model using the 2MASS stellar
   reference frame onto the GSC2.3 quasar positions. The clear points
   correspond to the quasars for which there are ERF radio positions.}
              \label{Fig11}%
    \end{figure*}

\section{The global orientation towards the ICRS}

In this section the points 7 and 8 of the Sect. 2.11 (data flow) are
derived.

Once the local correction is obtained, the next move is to tie
the preliminary frame just derived to the ICRF. As discussed in Sect. 1,
the ICRF alone is too sparse, and in particular the number of its bright
optical counterparts, to suffice for a detailed tying. Taking 
advantage of the fact that the long base radio interferometric observations
always combine data for several sources in different sessions, a radio position
net considerably larger than the ICRF was formed by adding the VLBA
calibrators and the VLA calibrators. In the case of the VLA, only
sources with the highest astrometric precision are included. The accurate,
ICRF representing, radio interferometric positions of the enlarged
radio sample sources (ERF) are then compared with their optical 
counterpart from the stellar based preliminary frame derived from
the local corrections. From the ICRF 718 sources are collected,
while from the VSC6 and the VLAC, 2,684 sources and 59 sources
are collected respectively. 

The enlarged radio frame so gathered thus contains 3,461 sources. They
are adequately represented in all of the three contributing catalogs 
as detailed in Table 3 (column $Nrad$). The sky distribution was shown in Fig. 4.
From the ERF sources, 2,263 have an optical counterpart in
at least one of the input catalogs.
The average minimum distance between the ERF sources is 187 arcmin, with a mode of 70 arcmin. 

The ERF radio positions are then compared with their optical 
counterpart from the stellar based preliminary frame derived by
the local corrections. 
Initially the comparison entails the derivation of global orientation
corrections about the equatorial coordinates standard triad, plus
a correction for a bias relatively to the equator. Following (Arias, Feissel \& Lestrade, 1988),
the canonical
equations relating the right ascension and declination offsets to 
the rotations (A1, A2, A3) about the equatorial axes (namely, X on the 
equator pointing to the conventional origin, Y on the equator 
perpendicular to X, and Z pointing to the conventional pole) are given by

\begin{eqnarray}
\Delta\alpha\,cos\delta &= A1\,sin\delta\,cos\alpha + A2\,sin\delta\,sin\alpha -
                         A3\,cos\delta
\label{eq:equation5}
\end{eqnarray}
\begin{eqnarray}
\Delta\delta &= -A1\,sin\alpha + A2\,cos\alpha + A4
\label{eq:equation6}
\end{eqnarray}\\

Equations (5) and (6) are solved independently as a measure to assess
the robustness of the corrections. In this case, the direction cosines
A1 and A2 agree when calculated either from $\Delta\alpha$cos$\delta$ or 
$\Delta\delta$ within 1 $\sigma$ in most cases. Figure 12 shows the good 
agreement on A1 and A2 as calculated from the right ascension residuals,
from the declination residuals, or from the combined residuals.
The equatorial bias (which is discussed in detail in the next section)
can also be independently derived as simply ${\Delta\delta}$ = $\it{A4}$. 
Also for A4, either the straight average or the adopted combined 
solution agree well.

   \begin{figure}
   \centering
   \includegraphics[width=9cm]{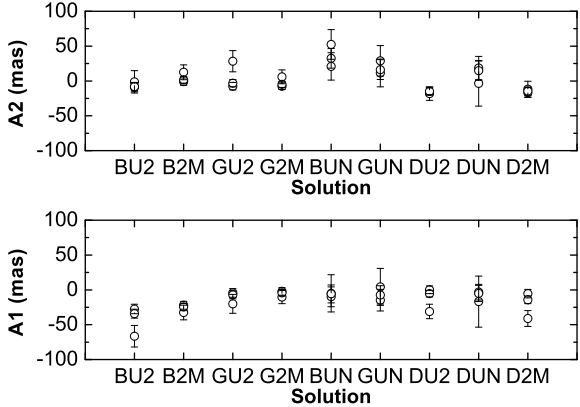}
   \caption{Coincidence of the A1 and A2 direction cosines as obtained 
   either from the $\Delta\alpha$cos$\delta$ residuals, from the $\Delta\delta$ 
   residuals, or from both. The value adopted was from the solution including 
   both the right ascension and declination residuals.}
              \label{Fig12}%
    \end{figure}

For the sake of gathering the largest number of equations of condition, 
the final values are derived by a combined matrix including the three 
rotation corrections, plus the representation of the equatorial bias.

\subsection{The equatorial bias}

Former investigations (Assafin et al., 2001) pointed out marked 
differences in the equatorial bias concerning the USNO A2.0 catalog.
The point of discontinuity seats around -20$^\circ$ of declination,
which is the boundary of the north and south plate surveys that 
contributed to that catalog. Although such marked discontinuity 
should not befall neither the B1.0 nor the GSC23 catalogs, they
mostly share the same constituent surveys with the USNO A2.0. 

In order to
analyze whether a discontinuity remains in the data, Eq. (6)
was solved for the catalogs B1.0 and GSC23, as locally corrected
by the UCAC2 and the 2MASS frames, adopting a single A4 
unknown, two A4 unknowns dividing the $\Delta\delta$ offsets at 
-20$^\circ$ of declination,
and eighteen A4 unknowns to fit the $\Delta\delta$ offsets within 
declination strips 10$^{\circ}$  wide. The plots in Fig. 13 show 
features that are more complex than could be described by one A4 standard offset. 
Nonetheless,
there is a coarse positive plateau southwards of declination -20$^\circ$, 
and values averaging near zero northwards. The standard deviation
in the residuals from the 18 strips solutions is indeed no smaller than that
of the -20$^\circ$ boundary, 2 A4 terms solution. This testifies of
an interplay that can be more adequately handled in the
next step where harmonic terms are introduced. On the other hand, 
the solution contemplating a single A4 term fares the worst by a
few mas, and is certain to provide the least appropriate description.

   \begin{figure*}
   \centering
   \includegraphics[width=18cm]{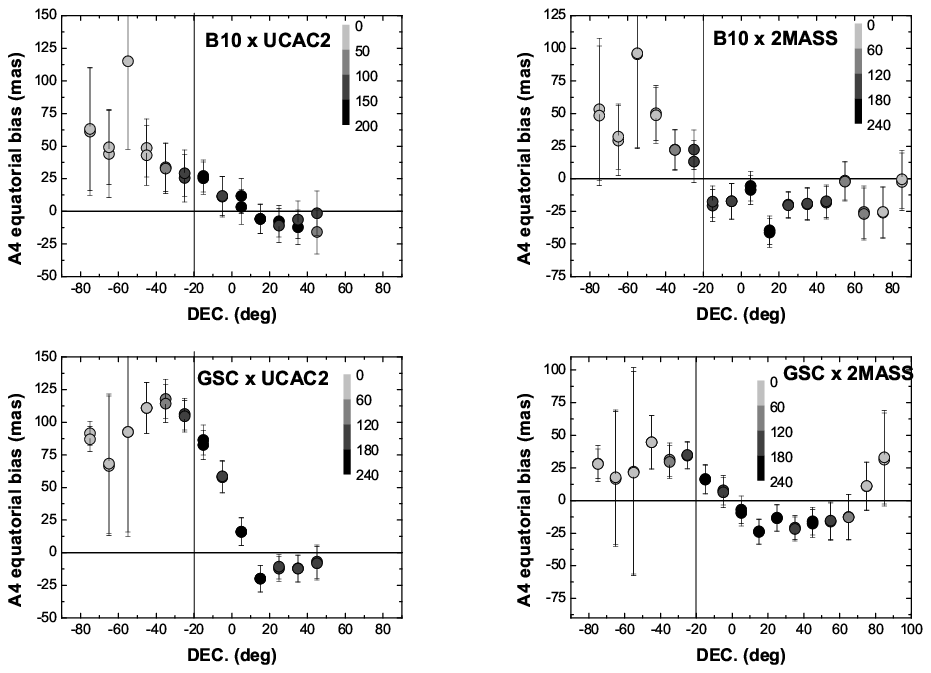}
   \caption{Declination dependency of the equatorial bias (A4).
   The plotted averages correspond to bins 10$^\circ$ wide in declination.
   The shaded codes indicate the number of ERF quasars within the bins.}
              \label{Fig13}%
    \end{figure*}

The verified A4 declination dependency, prompted to assess whether
a similar right ascension dependency is present. In this case, no
such dependency exists. 

\subsection{Rotation angles and equatorial bias}

The final values of the A1, A2, and A3 direction cosines for the
global orientation towards the ICRF, as well as the equatorial bias,
were found by adjusting the optical minus radio residual between
the positions evaluated from the local astrometric solution and
those given by the enlarged radio frame. The values are presented
in Table 4.

\begin{table*}[htb]
  \caption{Direction cosines relative to the ERF after the application of the local astrometric corrections$^{\mathrm{a}}$. }

  \label{tab:table4}
  \begin{center}
    \leavevmode
        \begin{tabular}[h]{lrccccc}
      \hline \\[-5pt]
      Solution & N & A1 & A2 & A3 & A4 & A5
       \\[+5pt]

      \hline \\[-5pt]

BU6 & 2616 & -33.7 $\pm$ 6.3 & -8.0 $\pm$ 6.1 & 50.8 $\pm$ 5.3 & 40.5 $\pm$ 10.7 & 2.6 $\pm$ 5.3 \\
BN6 & 244 & -5.5 $\pm$ 12.9 & 31.8 $\pm$ 14.1 & 123.0 $\pm$ 25.2  & 1.5 $\pm$ 12.8 & \\
BT6 & 3764 & -23.8 $\pm$ 5.1 & 1.5 $\pm$ 4.7 & 23.7 $\pm$ 4.6 & 27.5 $\pm$ 9.7 & -21.0 $\pm$ 4.3 \\
BU4 & 2522 & -37.5 $\pm$ 6.3 & -4.9 $\pm$ 6.2 & 48.6 $\pm$ 5.4 & 42.1 $\pm$ 10.7 & 0.0 $\pm$ 5.4 \\
BN4 & 240 & -0.4 $\pm$ 12.9 & 34.7 $\pm$ 14.1 & 118.9 $\pm$ 25.4 & -0.3 $\pm$ 12.8 & \\
BT4 & 3762 & -24.6 $\pm$ 5.1 & 1.4 $\pm$ 4.7 & 24.3 $\pm$ 4.6 & 24.7 $\pm$ 9.6 & -21.7 $\pm$ 4.3 \\
GU6 & 3238 & -7.0 $\pm$ 5.8 & -3.1 $\pm$ 5.3 & 30.4 $\pm$ 4.7 & 100.8 $\pm$ 9.3 & 15.2 $\pm$ 4.8 \\
GN6 & 300 & -7.2 $\pm$ 13.2 & 16.3 $\pm$ 14.2 & -43.4 $\pm$ 25.1 & -4.7 $\pm$ 13.1 & \\
GT6 & 3940 & -4.0 $\pm$ 4.7 & -4.9 $\pm$ 4.4 & 8.7 $\pm$ 4.3 & 30.2 $\pm$  8.7 & -6.1 $\pm$ 4.0 \\
GU4 & 3240 & -7.7 $\pm$ 5.8 & -2.5 $\pm$ 5.3 & 29.4 $\pm$ 4.7 & 99.1 $\pm$ 9.3 & 14.7 $\pm$ 4.8 \\
GN4 & 308 & -7.1 $\pm$ 13.1 & 15.7 $\pm$ 14.0 & -40.4 $\pm$ 24.7 & -6.8 $\pm$ 12.9 & \\
GT4 & 3938 & -3.7 $\pm$ 4.7 & -6.0 $\pm$ 4.4 & 8.7 $\pm$ 4.3 & 30.1 $\pm$ 8.7 & -6.3 $\pm$ 4.0 \\
SU6 & 480 & -5.3 $\pm$ 4.7 & -14.6 $\pm$ 4.8 & -4.8 $\pm$ 3.7 & 16.6 $\pm$ 4.1 & \\
SN6 & 74 & -5.0 $\pm$ 11.6 & 14.8 $\pm$ 13.9 & 32.0 $\pm$ 17.8 & 9.8 $\pm$ 13.1 & \\
ST6 & 600 & -13.9 $\pm$ 5.5 & -14.6 $\pm$ 6.0 & -5.1 $\pm$ 4.9 & 20.1 $\pm$ 5.3 & \\
SU4 & 478 & -5.5 $\pm$ 4.5 & -12.8 $\pm$ 4.6 & -5.2 $\pm$ 3.5 & 16.3 $\pm$ 4.0 & \\
SN4 & 74 & -6.2 $\pm$ 11.4 & 13.1 $\pm$ 13.8 & 33.2 $\pm$ 17.6 & 9.5 $\pm$ 13.0 & \\
ST4 & 600 & -14.3 $\pm$ 5.5 & -14.2 $\pm$ 6.0 & -4.7 $\pm$ 4.9 & 19.2 $\pm$  5.3 & \\

      \hline \\
      \end{tabular}
  \end{center}
  \begin{list}{}{}
  \item[$^{\mathrm{a}}$] Direction cosines A1, A2, and A3 from the comparison
  of the equatorial positions as found for each family of solutions and as
  given by the radio interferometry for the ERF objects. 
  The equatorial bias A4 (north of declination -20$^\circ$) and 
  A5 (south of declination -20$^\circ$), when applicable, are also presented.
  The solutions nomenclature is formed by one letter
           designating the quasar position input catalog, followed
           by one letter designating the stellar reference frame, 
           and by the codes $6$ and $4$ informing whether the algorithm
           is of the six or four parameter models.
            N is the number of equations used to derive the direction cosines
            and equatorial biases. All angular values are in milli-arcsec.
  \end{list}
\end{table*}

For the B1.0 and GSC2.3 northernmost sources, for
which the local correction was made using the UCACN, separate
values were derived for the rotation angle and the (northern only)
equatorial bias. In what concerns the DR5 northernmost sources,
though the number of UCACN treated sources is small, the precisions are 
high and so a complete solution including one equatorial bias term
was sufficient. At large, for the DR5 the A1 and A2 values,
although significant and representing true corrections to be applied
to the quasar equatorial coordinates, must be interpreted with caution
in terms of what they would represent for the DR5 frame orientation, because
of the concentration of the DR5 sources in the direction of the right
ascension of 12h.

\section{The removal of zonal bias}

In this section the points 9 and 10 of the Sect. 2.11 (data flow) are
derived.

\subsection{Harmonic functions}

\begin{table*}[htb]
  \caption{ Number of significant harmonic terms in the three rounds
  of testing$^{\mathrm{a}}$.}

  \label{tab:table5}
  \begin{center}
    \leavevmode
        \begin{tabular}[h]{lcccccc}
      \hline \\[-5pt]
      Solution &  & R.A. &  &  & DEC. &  \\
       \hline \\
       & 1$^{st}$ round & 2$^{nd}$ round & 3$^{rd}$ round & 1$^{st}$ round &
 2$^{nd}$ round & 3$^{rd}$ round \\
      \hline \\[-5pt]

BU6 & 21 & 6 & 6 & ~2 & 1 & 1 \\
BN6 & ~0 & ~0 & ~0 & ~0 & 0 & 0 \\
BT6 & 17 & 9 & 9 & 17 & 5 & 5 \\
BU4 & 23 & 7 & 7 & ~0 & ~0 & ~0 \\
BN4 & ~0 & ~0 & ~0 & ~0 & 0 & 0 \\
BT4 & 17 & 9 & 9 & 17 & 5 & 5 \\
GU6 & 15 & 6 & 6 & ~8 & 3 & 3 \\
GN6 & ~0 & ~0 & ~0 & ~2 & 1 & 1 \\
GT6 & ~9 & 4 & 4 & ~5 & 2 & 2 \\
GU4 & 15 & 6 & 6 & ~8 & 2 & 2 \\
GN4 & ~0 & ~0 & ~0 & ~2 & 1 & 1 \\
GT4 & ~9 & 4 & 4 & ~5 & 2 & 2 \\
SU6 & ~0 & 0 & 0 & ~0 & 0 & 0 \\
SN6 & ~0 & 0 & 0 & ~0 & 0 & 0 \\
ST6 & ~1 & 1 & 1 & ~0 & 0 & 0 \\
SU4 & ~0 & 0 & 0 & ~0 & 0 & 0 \\
SN4 & ~0 & 0 & 0 & ~0 & 0 & 0 \\
ST4 & ~1 & 1 & 1 & ~0 & 0 & 0 \\

      \hline \\
      \end{tabular}
  \end{center}
 \begin{list}{}{}
 \item[$^{\mathrm{a}}$] Harmonic terms surviving in the three rounds
  of testing for the most significant ones. The three rounds are explained
  in the text, and the third round corresponds to the adopted harmonic fit.
  All procedure was done independently for the right ascension and
  declination residuals between the solutions and the ERF radio interferometry
  positions.
  The solutions nomenclature is formed by one letter
           designating the quasar position input catalog, followed
           by one letter designating the stellar reference frame,
           and by the codes $6$ and $4$ informing whether the algorithm
           is of the six or four parameter model.
\end{list}
\end{table*}

At this point, the optical frame was homogenized by the local
corrections, and was globally aligned with the ICRF. The final point
is to sweep its axes to check for any remaining zonal modulations. This
was done for the right ascension, declination, and R magnitude dimensions.
In a sense this is equivalent to straighten up the axes from waves
of corrugation either trapped in the stellar catalogs themselves,
or that were present in the initial deep catalogs and could not
be completely removed by the local corrections. 

The problem, the quantity of assessing points, the sky coverage, the
dimensions of dependencies, and the relationship from the less
precise, corrugated frame to the more precise, correcting one, are all
the same to the case treated by Schwan (1988) when deriving zonal
corrections to the FK5 construction. Thus, the same treatment by spherical
harmonics (Brosche, 1966) is used here. The right ascension and declination corrections
are expressed by the linear combination of the terms of a series,
in which the $cpnml-th$ element is given by

\begin{eqnarray}
\Delta\,c &= Ccpnml\,Hp(R')\,Ln(sin\delta)\,Fml(\alpha)
\label{eq:equation7}
\end{eqnarray} 

where ${\Delta c}$ is one or the other equatorial coordinates offsets,
since independent spherical harmonic series are derived for each of
them; $\it{Ccpnml}$ is the $\it{cprml-th}$ coefficient to be derived, 
for the $\it{c}$ equatorial coordinate series; $Hp(R')$ is the 
Hermite polynomial of degree $p$, function of the R magnitude normalized
by the mean and standard deviation for the quasars; $Ln(sin\delta)$ 
is the Legendre polynomial of degree $n$, function of $sin\delta$; and
$Fml(\alpha)$ is the Fourier term of order $m$ on $\alpha$ and sign $l$.

The offsets are produced by comparing the equatorial coordinates
from the previous step (global rotation) with the VLBI positions from
the ERF. That is, 2.263 pairs are available. Again, this is similar to
the quantity available for the Schwan's (1988) application. Therefore
terms up to very higher order can be derived. However, the analysis from
the earlier steps advises to some caution in defining the actually significant
order up which to halt. The distribution in right ascension of the
offsets from the original deep catalogs hints for a 2$\pi$ term and
smaller ripples (cf. Figs. 5-7), while the analysis of global rotation
coefficients has shown no $\alpha$ dependence. In contrast, the analysis
of the equatorial bias displays a $\delta$ dependence, coherent in
bins of 10$^\circ$ (Fig. 13). Magnitude dependences are believed to be possible
within the large quasar catalogs. They could to some extent
be introduced by the local solution itself, because of the magnitude
difference between stars and quasars. The analysis made in Sect. 3.3 
points out to such dependences but of feeble amplitude. All
accounted for, and in order to have representative quantities of ERF
objects with which to compute the coefficients of each term, it was 
adopted to extend to 9$^{th}$ order the Lagrange ($sin\delta$)
polynomials and the Fourier ($\alpha$) functions. The Hermite (R magnitude)
polynomials were extended to 2$^{nd}$ order, but just to the 5$^{th}$ 
order of the other two terms. Given the poor sky distribution, no declination
terms were used for catalog combinations including the UCACN frame,
nor right ascension terms for catalog combinations including the DR5 quasar
list. All the above choices were made to retain only truly
significant terms derived from the ERF comparison, such that they 
could be confidently applied to the full set of quasars.

Since by definition the spherical harmonics are orthogonally independent, their
statistical significance can be tested individually. Each term was initially 
tested separately. Only those statistically significant at 3$\sigma$ with
coefficients not smaller than 1 mas were retained. Once more, this taxes
heavier on the high order polynomials, Next, starting from the lower orders
and up, the corrections brought by each term are removed from the 
equatorial offsets, and the following term is individually tested. The
same criterion of 3$\sigma$ significance and coefficient not smaller than
1 mas is applied to retain the terms. Actually, at this round the uppest
order Hermite magnitude polynomials, when convolved with the higher 
Legendre (declination) and Fourier (right ascension) orders, are retained 
only if the coefficient is not smaller than 5 mas. The terms surviving these
two rounds are added by a linear combination to separately adjust the 
right ascension and declination offsets by least squares. Table 5 reports
the number of terms surviving the rounds. Notice the important decreasing rate
from the first to the second round, which justifies the rounds scheme. 
On the other hand, there is no dropping from the second round (one by one regimen)
to the third round (least squares fit of all validated terms together),
what shows that they are truly statistically significant.

Table 6 reports the coefficients derived at the final adjustment for
each family of quasar catalog and stellar frame. As expected, the coefficients
do not vary from the second round to the final adjustment, indicating
that no secondary harmonic survived. There is only one significant term
including the DR5. It relates to the 2MASS local correction and depends
solely on declination (3$^{rd}$ order). This reflects the high quality
of the DR5 positions and indicates that the local correction did not induce further
derail upon the input quasar catalogs. Concerning the UCACN there is also 
only one significant term. It relates to the GSC2.3 quasar catalog and is
of high right ascension order (8$^{th}$). The lack of UCACN harmonic terms 
reflects the confinement of the sky region where it was applied. The positions
originated from the B1.0 required 42 harmonic terms (31 for the right ascension
offsets), and those from the GSC23 32 harmonic terms (20 for the right ascension
offsets). Considering the stellar reference frames, the UCAC2 required
32 harmonic terms (25 for the right ascension offsets), while the 2MASS
required 42 harmonic terms (28 for the right ascension offsets).
In all cases, the right ascension offsets required more harmonic
terms. In contrast, the Ln(sin$\delta$) dominated terms in general correspond to
larger coefficients, which worked to correct the deviations detected at the 
A4 equatorial
bias analysis. The magnitude terms appear only to the 1$^{st}$ order, in
18 B1.0 related terms and 10 GSC23 related terms. Thus, magnitude terms
are presented as the catalogs´ authors cautioned, but they are less
important than the zonal terms. Figure 14 shows the histograms of the 
harmonic terms coefficients and associated errors. The errors resemble
well a Poisson distribution. The coefficients peak around -6 mas 
for the right ascension offsets, and around +6 mas for the declination
offsets.

\begin{table*}[htb]
  \caption{Values of the coefficients of the adjusted harmonic functions$^{\mathrm{a}}$.
   .}

  \label{tab:table6}
  \begin{center}
    \leavevmode
        \begin{tabular}[h]{llcccc}
      \hline \\[-5pt]
Solution & Offset & Hp(R') & Ln(sin$\delta$) & Fml($\alpha$) & Coefficient \\   
      \hline \\[-5pt]

BU6 & $\Delta\alpha$cos$\delta$ &  0 & 0 & 1~~~$-$1 & $+$18.7 $\pm$ 5.5 \\
BU6 & $\Delta\alpha$cos$\delta$ &  0 & 0 & 1~~~$+$1 & $+$41.7 $\pm$ 5.2 \\
BU6 & $\Delta\alpha$cos$\delta$ &  0 & 4 & 0~~~~0 & $-$21.0 $\pm$ 6.0 \\
BU6 & $\Delta\alpha$cos$\delta$ &  1 & 0 & 0~~~~0 & $-$~5.5 $\pm$ 2.7 \\
BU6 & $\Delta\alpha$cos$\delta$ &  1 & 3 & 0~~~~0 & $+$13.1 $\pm$ 2.9 \\
BU6 & $\Delta\alpha$cos$\delta$ &  1 & 4 & 1~~~$-$1 & $+$~8.9 $\pm$ 2.8 \\
BU6 & $\Delta\delta$            &  0 & 8 & 3~~~$+$1 & $-$20.2 $\pm$ 6.0 \\
\\
BN6 & $\Delta\alpha$cos$\delta$ & - & - & - & - \\
BN6 & $\Delta\delta$ & - & - & - & - \\
\\
BT6 & $\Delta\alpha$cos$\delta$ &  0 & 0 & 1~~~$-$1    &   $+$16.6 $\pm$    4.7 \\
BT6 & $\Delta\alpha$cos$\delta$ &  0 & 0 & 1~~~$+$1   &   $+$38.0  $\pm$   4.2 \\
BT6 & $\Delta\alpha$cos$\delta$ &  0 & 3 & 1~~~$+$1   &   $+$12.3  $\pm$   4.4 \\
BT6 & $\Delta\alpha$cos$\delta$ &  0 & 4 & 0~~~~0   &  $-$15.1  $\pm$   4.6 \\
BT6 & $\Delta\alpha$cos$\delta$ &  0 & 4 & 4~~~$-$1   &  $-$18.2  $\pm$   4.5 \\
BT6 & $\Delta\alpha$cos$\delta$ &  1 & 0 & 0~~~~0   &   $-$~8.7  $\pm$   2.1 \\
BT6 & $\Delta\alpha$cos$\delta$ &  1 & 3 & 0~~~~0   &   $+$~7.5   $\pm$  2.1 \\
BT6 & $\Delta\alpha$cos$\delta$ &  1 & 5 & 0~~~~0   &   $-$~7.0  $\pm$   2.1 \\
BT6 & $\Delta\alpha$cos$\delta$ &  1 & 5 & 1~~~$+$1   &    $+$~7.2  $\pm$   2.2 \\
BT6 & $\Delta\delta$ &  0 & 6 & 1~~~$-$1   &  $+$14.9  $\pm$   4.8 \\
BT6 & $\Delta\delta$ &  0 & 8 & 3~~~$+$1   &  $-$17.3 $\pm$    4.5 \\
BT6 & $\Delta\delta$ &  0 & 9 & 6~~~$+$1   &  $-$14.2 $\pm$    4.6 \\
BT6 & $\Delta\delta$ &  1 & 3 & 1~~~$+$1   &    $+$~6.9 $\pm$    2.0 \\
BT6 & $\Delta\delta$ &  1 & 3 & 3~~~$-$1   &    $+$~7.2 $\pm$    2.1 \\
\\
GU6 & $\Delta\alpha$cos$\delta$ & 0 & 0 & 1~~~$+$ & $+$18.0 $\pm$ 4.7 \\
GU6 & $\Delta\alpha$cos$\delta$ & 0 & 1 & 0~~~~0 & $+$64.2 $\pm$ 9.0 \\
GU6 & $\Delta\alpha$cos$\delta$ & 0 & 3 & 0~~~~0 & $-$36.3 $\pm$ 8.5 \\
GU6 & $\Delta\alpha$cos$\delta$ & 0 & 5 & 0~~~~0 & $+$22.0 $\pm$ 6.1 \\
GU6 & $\Delta\alpha$cos$\delta$ & 0 & 9 & 0~~~~0 & $+$19.9 $\pm$ 5.4 \\
GU6 & $\Delta\alpha$cos$\delta$ & 1 & 0 & 1~~~$+$ & $-$~7.1 $\pm$ 2.4 \\
GU6 & $\Delta\delta$ &  0 & 1 & 0~~~~0 & $-$23.0 $\pm$ 6.1 \\
GU6 & $\Delta\delta$ &  0 & 2 & 4~~~$+$1 & $-$18.6 $\pm$ 5.7 \\
GU6 & $\Delta\delta$ &  1 & 0 & 0~~~~0 & $+$~7.3 $\pm$ 2.3 \\
\\
GN6 & $\Delta\alpha$cos$\delta$ & - & - & - & - \\
GN6 & $\Delta\delta$ & 0 & 0 & 9~~~$-$1 & $-$33.9 $\pm$ 1,1 \\
\\
GT6 & $\Delta\alpha$cos$\delta$ &  0 & 0 & 1~~~$+$1 & $+$22.4 $\pm$ 4.0 \\
GT6 & $\Delta\alpha$cos$\delta$ &  0 & 1 & 0~~~~0 & $+$26.0 $\pm$ 4.6 \\
GT6 & $\Delta\alpha$cos$\delta$ &  0 & 3 & 0~~~~0 & $-$21.0 $\pm$ 4.4 \\
GT6 & $\Delta\alpha$cos$\delta$ &  1 & 1 & 1~~~$-$1 & $-$~7.9 $\pm$ 2.2 \\
GT6 & $\Delta\delta$ &  1 & 0 & 0~~~~0 & $+$~6.8 $\pm$ 2.1 \\
GT6 & $\Delta\delta$ &  1 & 2 & 4~~~$+$1 & $+$~7.0 $\pm$ 2.1 \\
\\
DU6 & $\Delta\alpha$cos$\delta$ & - & - & - & - \\
DU6 & $\Delta\delta$ & - & - & - & - \\
\\
DN6 & $\Delta\alpha$cos$\delta$ & - & - & - & - \\
DN6 & $\Delta\delta$ & - & - & - & - \\
\\
DT6 & $\Delta\alpha$cos$\delta$ & 0 & 4 & 0~~~~0 & $-$14.8 $\pm$ 3.8 \\
DT6 & $\Delta\delta$ & - & - & - & - \\
\\

      \hline \\
      \end{tabular}
  \end{center}
  \begin{list}{}{}
\item[$^{\mathrm{a}}$] As the results from the six constants and four constants local astrometreic
     solutions are quite alike, only the results from the six constants solutions
     are displayed. The code for the solutions are as in the previous tables 
   and the coefficients are in milli-arcsec.
\end{list}

\end{table*}

   \begin{figure}
   \centering
   \includegraphics[width=9cm]{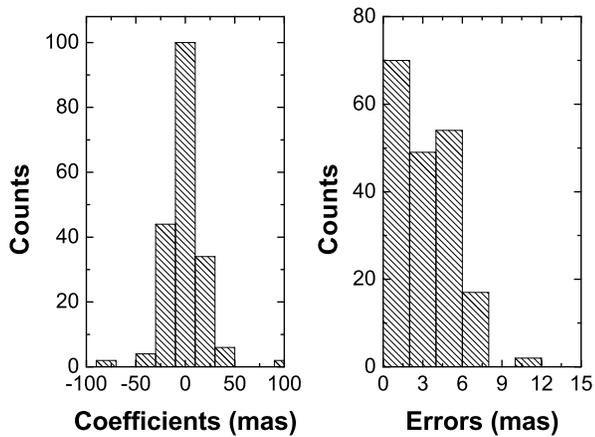}
   \caption{Distribution of the coefficients and associated errors of the
   harmonic functions fitting the $\Delta\alpha$cos$\delta$ and the $\Delta\delta$
   residuals. All significant terms (first round - see text) are counted here.}
              \label{Fig14}%
    \end{figure}

\subsection{Local inhomogeneities}

Figures 15 and 16 (open symbols) show that after the correction by harmonic series, the
optical minus radio residuals are equal to zero within the statistical
significance of 3 $\times$ standard error, that is without being attached
to the scatter of the optical position determinations. This is shown for
the pairs combining the B1.0 (Fig. 15) and GSC (Fig. 16) quasar input lists, 
and the UCAC2 (squares) and 2MASS (circles) local stellar frames, which are 
the largest and have the most 
homogenous sky coverage of the combinations used here.

Notwithstanding this, some clumping is seen, without a clear pattern over the sky, 
that can be dealt with further. The ERF sky density is such that the mean
distance from any given quasar to the closest ERF VLBI position is 1.7$^{\circ}$.
And, on average, there are 10 ERF VLBI positions within a radius of 4.4$^{\circ}$
around any given quasar, the largest distance being on average 6.3$^{\circ}$. 
Under such conditions for each of the combinations of quasar list and
stellar reference frame, hence for every quasar a correction for the
local inhomogeneities towards the
VLBI position can be obtained from the average of 10 ERF quasars.
Within the ERF quasars radius, objects are removed one by one if
both their right ascension and declination offsets are larger than 2$\sigma$.
On average, 1.3 ERF quasars are removed, what shows the robustness of
the local inhomegeneities corrections. This procedure is similar to that
of overlapping circles (Taff et al., 1990), but no overlap is actually required
since the autocorrelation between adjacent corrections fades into a
characteristic value at a distance of 20$^{\circ}$. Figure 17 illustrates 
the weakness of the autocorrelations
by taking each quasar in turn as a pole source, and plotting the standard
deviation in the inhomogeinity correction within growing distance rings.
For nearby sources, i.e., very small rings, the inhomogeneities are much
the same, and the standard deviation is accordingly small. Rapidly though,
before reaching the distance of 20$^{\circ}$, the standard deviation already
reaches the ceiling of scatter characteristic of each family of input catalog
and stellar reference frame solutions.

By determining in this way the correction for the local inhomogeneities,
it is seen in Figs. 15 and 16 (filled symbols) that the clumps are clearly minimized. 
After applying the corrections by harmonic functions 
the optical minus radio offsets are typically null within 1.3$\sigma$, what is
reduced further to within 0.7$\sigma$ after applying the complementary local inhomogeneities
corrections .

   \begin{figure*}
   \centering
   \includegraphics[width=18cm]{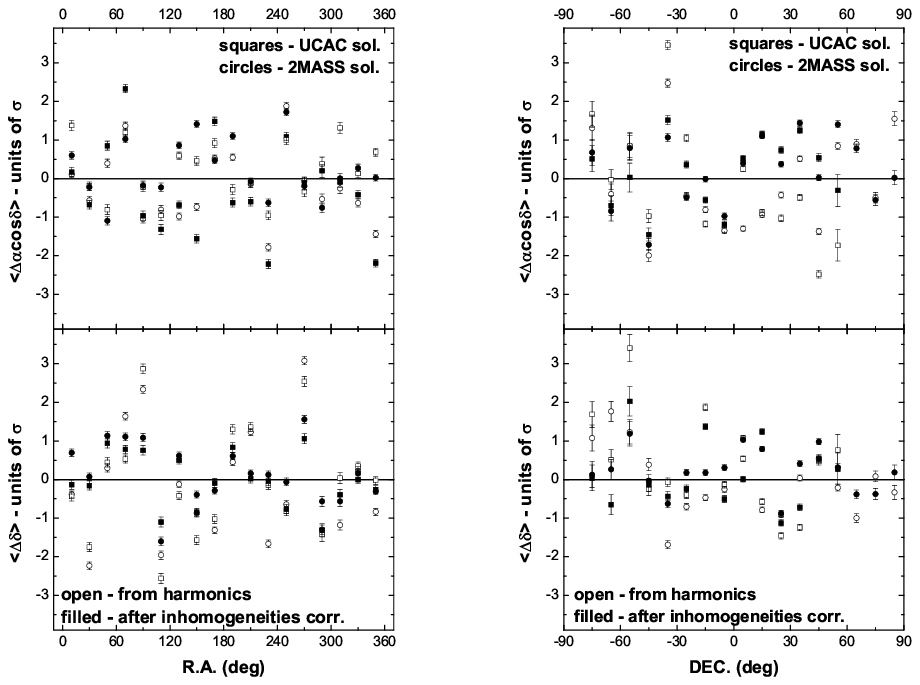}
   \caption{Averaged $\Delta\alpha$cos$\delta$ and $\Delta\delta$ optical
   (from the USNO B1.0) minus radio (from the ERF) residuals in units of 
   their standard deviations. Open symbols refer to the values corrected by 
   the harmonic functions. Filled symbols refer to the values further 
   redressed by the local inhomogeneities corrections. In both cases the 
   averages are non significant below 3$\sigma$. The actual significance
   averages are 1.2$\sigma$ after the harmonics treatment and 0.8$\sigma$ after
   the complementary local inhomogeneities corrections.} 
   
              \label{Fig15}%
    \end{figure*}

   \begin{figure*}
   \centering
   \includegraphics[width=18cm]{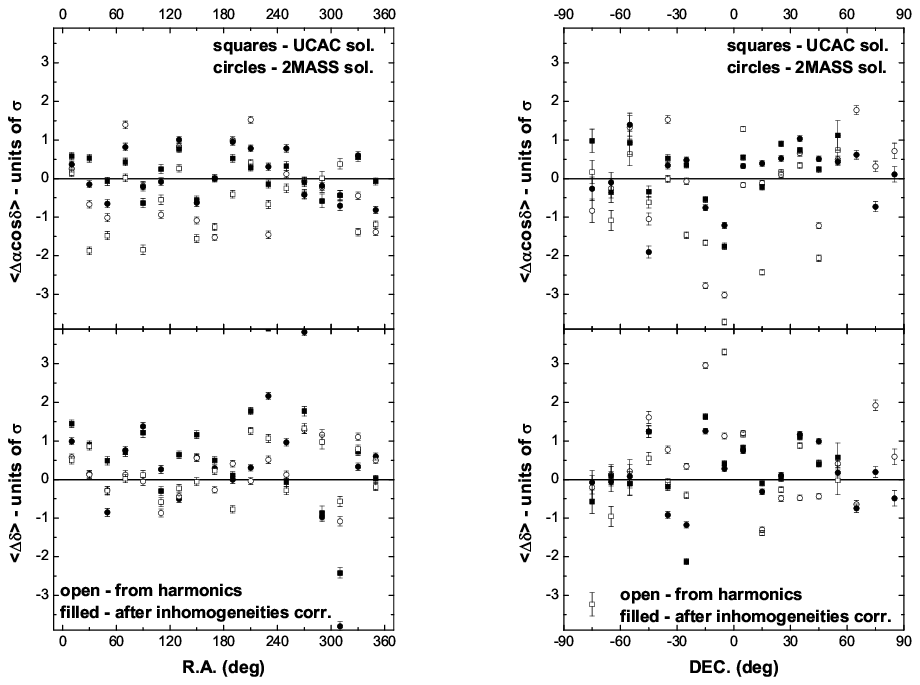}
   \caption{Averaged $\Delta\alpha$cos$\delta$ and $\Delta\delta$ optical
   (from the GSC2.3) minus radio (from the ERF) residuals in units of 
   their standard deviations. Open symbols refer to the values corrected by 
   the harmonic functions. Filled symbols refer to the values further 
   redressed by the local inhomogeneities corrections. In both cases the 
   averages are non significant below 3$\sigma$. The actual significance
   averages are 1.3$\sigma$ after the harmonics treatment and 0.7$\sigma$ after
   the complementary local inhomogeneities corrections..}
              \label{Fig16}%
    \end{figure*}

   \begin{figure}
   \centering
   \includegraphics[width=9cm]{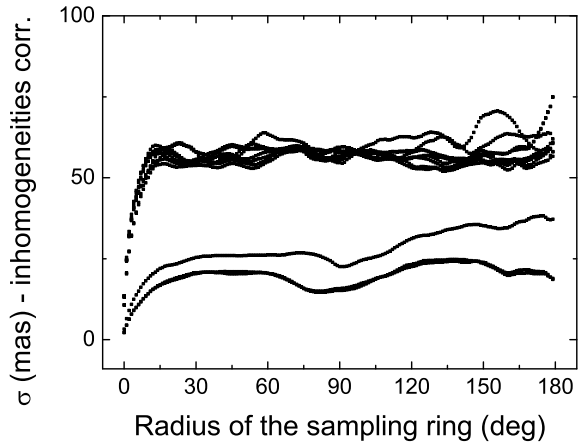}
   \caption{Fast fading of the autocorrelation of the local inhomogeneities
   correction. In the figure is plotted the standard deviation of the corrections 
   in samples taken in rings of increasing radius. The upper lines 
   correspond to the USNO B1.0 and GSC2.3 families of solutions, while the
   lower lines correspond the SDSS DR5 families of solutions.}
              \label{Fig17}%
    \end{figure}

\section{The LQRF catalog}

The LQRF catalog right ascension and declination equatorial coordinates 
are calculated from the values obtained in the final step of the data
treatment (removal of zonal 
bias) by a weighted average. In the average, for each quasar,  
the solutions from the three input lists (B1.0, GSC23, and DR5) and the
three stellar reference frames (UCAC2, UCACN, and 2MASS) are combined. 
The weights
are the inverse of the square root sum of the internal and external errors
at each of the correction steps plus the formal errors of each entry 
quasar input list. In order though to not carry too high or
too low a weight for any given combination, for each of them separately
the errors have been assigned by quartiles. It is worth noticing that
solutions derived with different weighting schemes or even without any weighting
agree within 10 mas. This indicates that the goal of homogenizing the 
different inputs was accomplished.

Since the final positions originate from the average of multiple combinations,
attaching the final error of them to the squared sum of the contributing errors
would in fact privilege the sources with fewer solutions obtained. Likewise,
if attaching the error to the dispersion in the multiple contributing
solutions. It is more realistic to derive the final error in the equatorial
coordinates of each source from the external comparison of the LQRF position
with the radio position. Since in the construction of the LQRF the three
correction steps focused on the achievement of an homogeneous frame, this is done by taking 
the average of the LQRF minus radio positions for the 10 closest ERF sources
around each quasar. As seen before, this implies a typical radius of 4.4$^{\circ}$
around any LQRF quasar. Figure 18 presents the histograms of the errors in
the coordinates. 
The distributions can be reckoned to Poisson distributions peaking at
   139 mas in right ascension and at 130 mas in declination.

   \begin{figure}
   \centering
   \includegraphics[width=9cm]{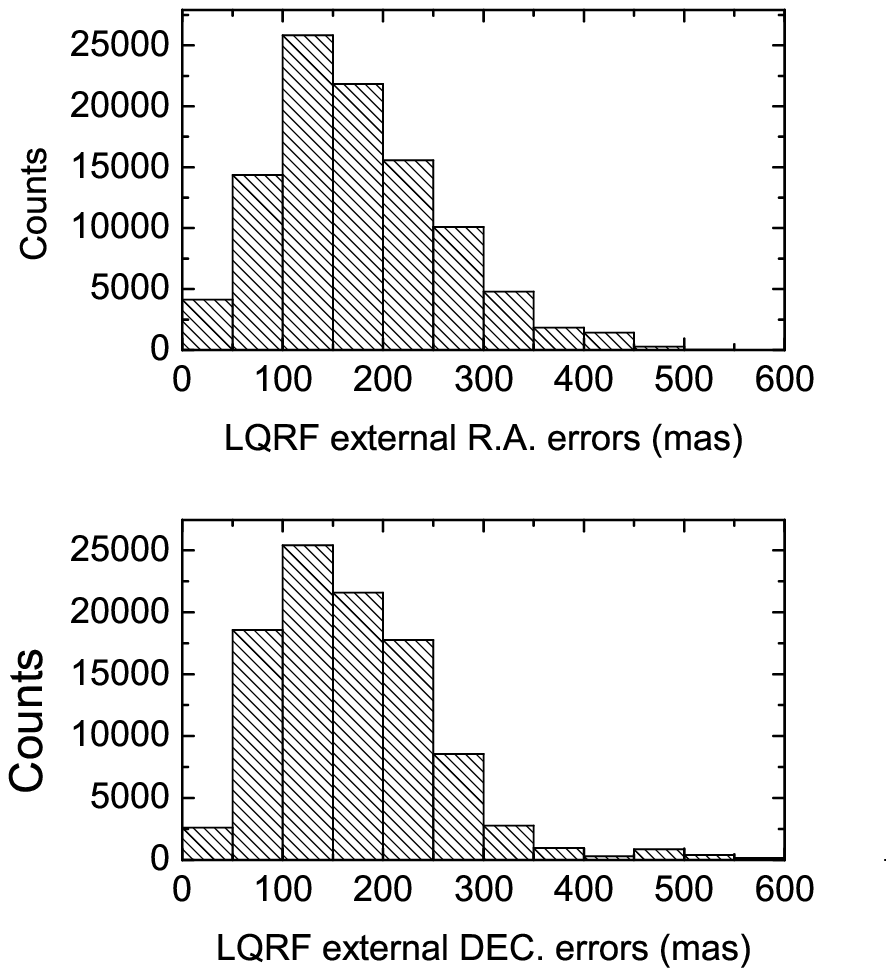}
 \caption{Distribution of the external errors assigned to the right
   ascension and declination J2000 positions of the LQRF sources. The
   distributions can be reckoned to Poisson distributions peaking at 
   139 mas on right ascension and at 130 mas on declination.}
              \label{Fig18}%
    \end{figure}

The LQRF contains 100,165 sources. On average, every source has a neighbor
within 10 arcmin. Figure 19 shows the sky distribution counts within squares 
of 10$^{\circ}$. Empty boxes are found only in the Galactic plane towards the 
southern hemisphere. Figure 20 presents a vector map of the offsets versus the VLBI 
positions, in bins of side 30$^{\circ}$ and at least 5 quasars. The mean offset
is 32.7 mas and their distribution appears as random, with the largest offsets
corresponding to bins of only 5 or 6 quasars.

   \begin{figure}
   \centering
   \includegraphics[width=9cm]{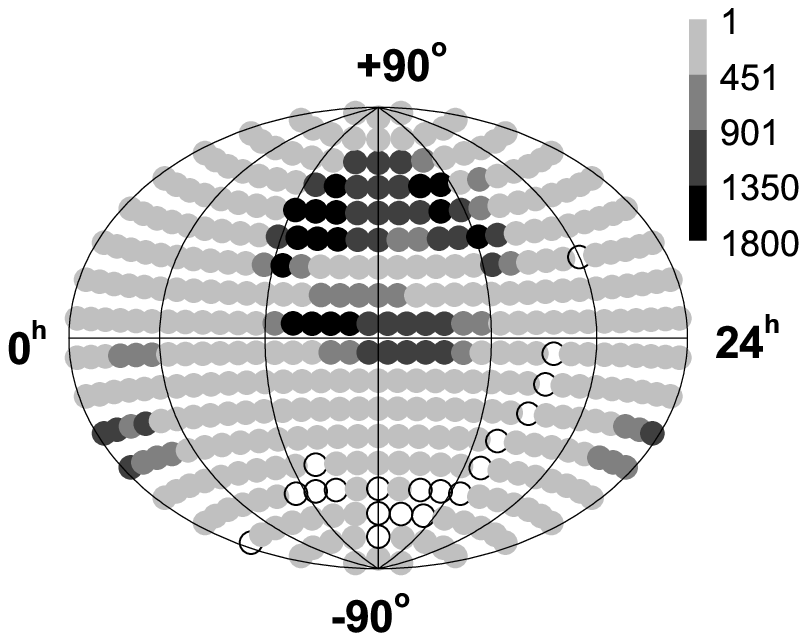}
   \caption{Sky density of the LQRF. The counts are in bins of 10$^{\circ}$ .
   The void regions shown by white circles are in the galactic plane. The
   densest patches lie on the SDSS region.}
              \label{Fig19}%
    \end{figure}

   \begin{figure}
   \centering
   \includegraphics[width=9cm]{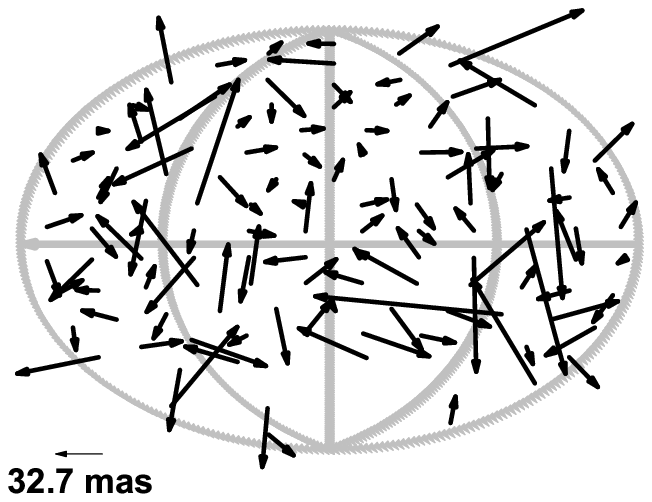}
   \caption{Vectorial distribution of the systematic deviations (north up, 
   east right) of the LQRF to the ICRF within bins of 10$^{\circ}$ . The average
   value is 32.7 mas, graphically represented by the off map arrow. All large values
   however befall to bins with 5 or 6 sources only.}
              \label{Fig20}%
    \end{figure}

Figures 21 to 24 show the optical minus radio coordinates offsets. The
comparison with the equivalent Figs. 5-7 illustrates clearly the 
evenness of the ICRF representation produced by the LQRF positions.

   \begin{figure}
   \centering
   \includegraphics[width=9cm]{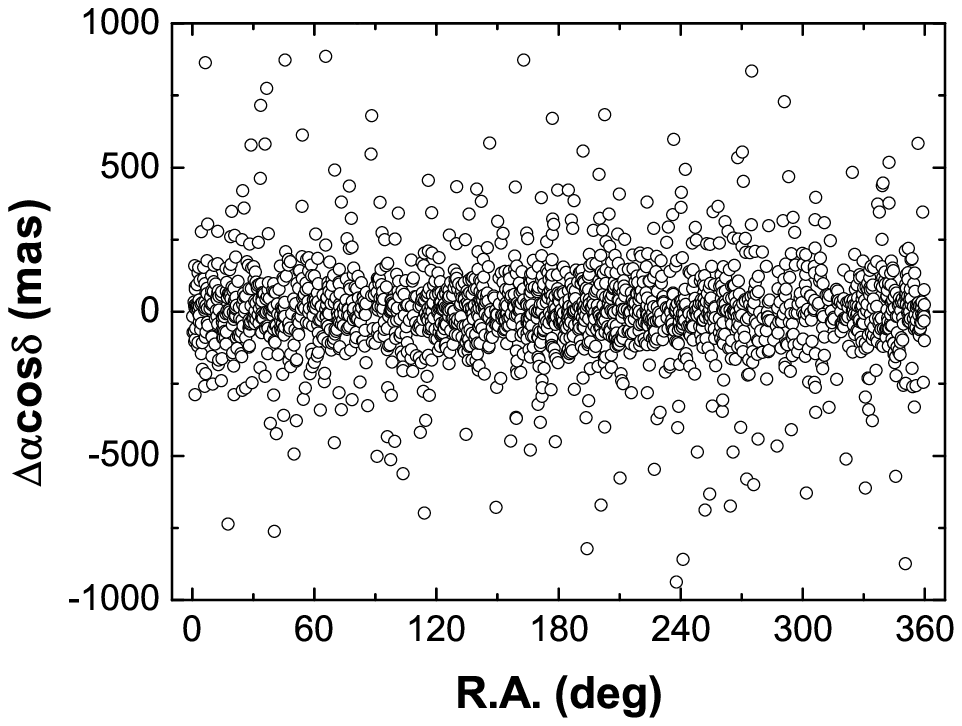}
   \caption{Distribution on right ascension of the
   $\Delta\alpha$cos$\delta$ LQRF minus ICRF residuals. There is no 
   systematics above 4 mas and the median is 108.3 mas.}
              \label{Fig21}%
    \end{figure}

   \begin{figure}
   \centering
   \includegraphics[width=9cm]{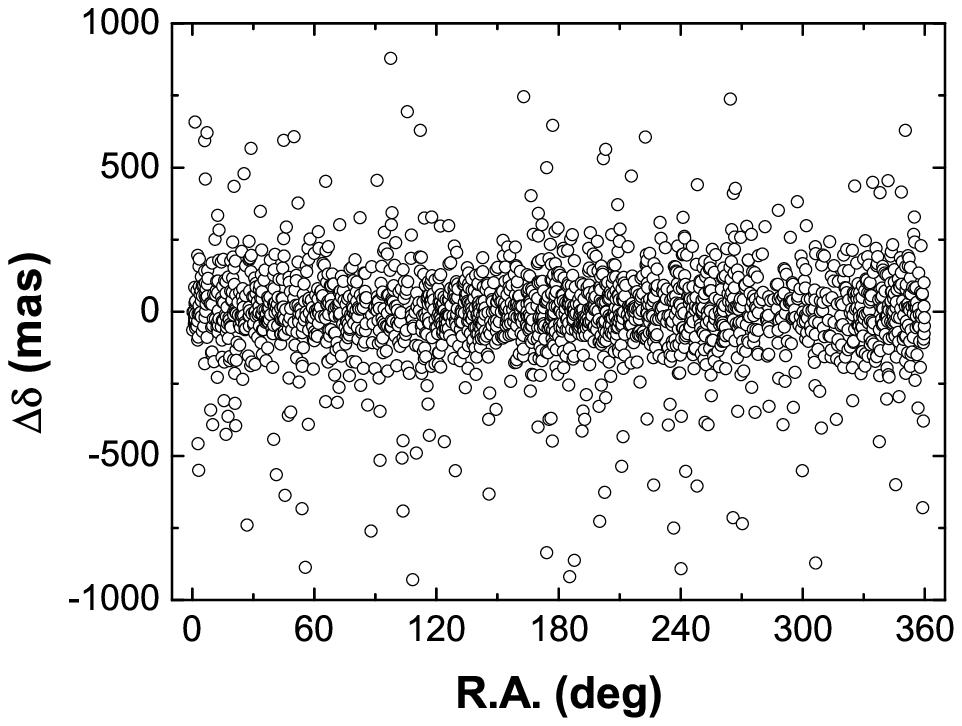}
   \caption{Distribution on right ascension of the
   $\Delta\delta$ LQRF minus ICRF residuals. There is no 
   systematics above 4 mas and the median is 105.3 mas.}
              \label{Fig22}%
    \end{figure}

   \begin{figure}
   \centering
   \includegraphics[width=9cm]{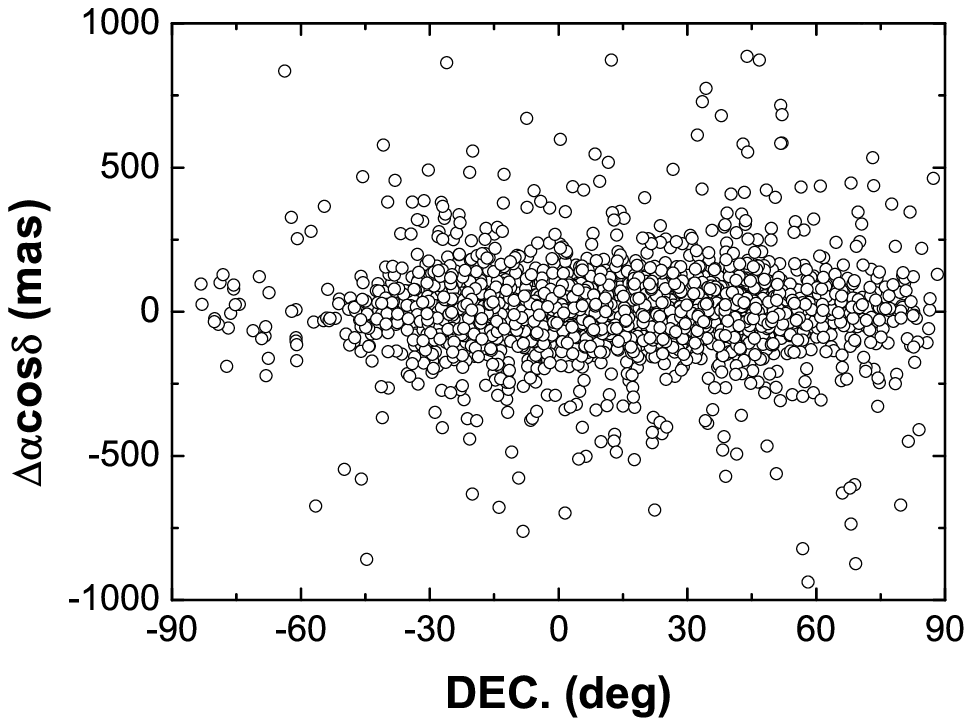}
   \caption{Distribution on declination of the
   $\Delta\alpha$cos$\delta$ LQRF minus ICRF residuals. There is no 
   systematics above 4 mas and the median is 108.3 mas.}
              \label{Fig23}%
    \end{figure}

   \begin{figure}
   \centering
   \includegraphics[width=9cm]{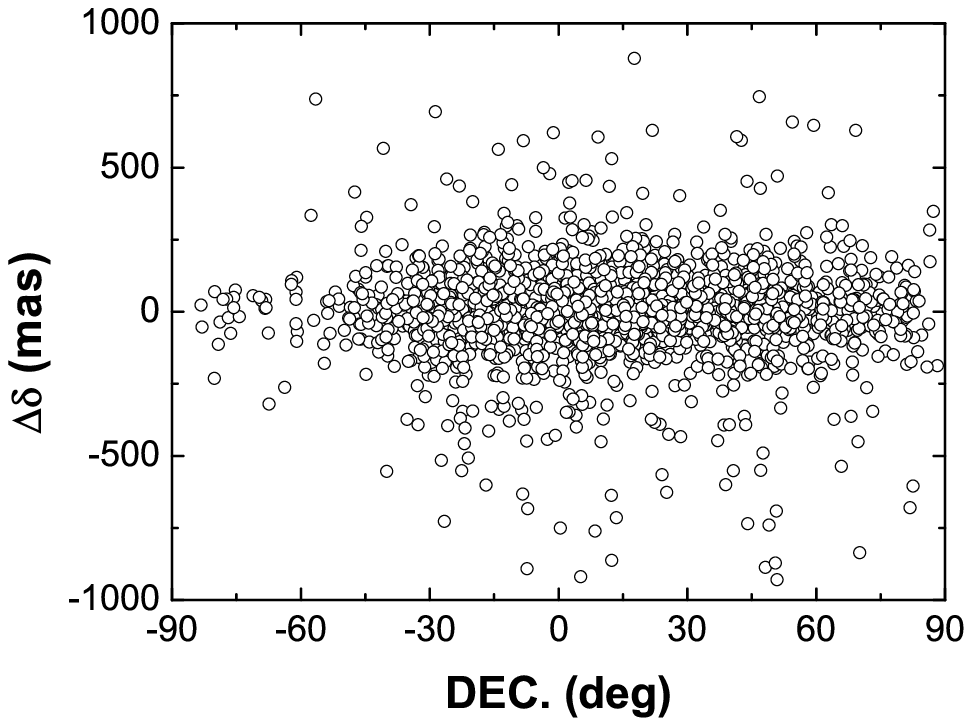}
   \caption{Distribution on declination of the
   $\Delta\delta$ LQRF minus ICRF residuals. There is no 
   systematics above 4 mas and the median is 105.3 mas.}
              \label{Fig24}%
    \end{figure}

The LQRF catalog is available through CDS access (in electronic form
via anonymous ftp to cdsarc.u-strasbg.fr (130.79.128.5)
or via http://cdsweb.u-strasbg.fr/cgi-bin/VizieR?-source=I/313). 
An extract of its first page appears in Fig. 25. Its columns contain 

 \begin{enumerate}
 
\item the J2000 right ascension (h,m,s);

\item the J2000 declination (d,',");

\item the external right ascension error ($\times$ cos$\delta$) (mas);

\item the external declination error (mas);

\item the number of radio interferometry neighbor positions used to evaluate 
the external errors;

\item the R magnitude, taken from the LQAC;

\item the redshift, taken from the LQAC (blank if not defined);

\item the right ascension offsets to the radio interferometry position (blank
if there is no radio position) ($\times$ cos$\delta$) (mas);

\item the declination offsets to the radio interferometry position (blank
if there is no radio position) (mas);

\item the J2000 radio interferometry right ascension (h,m,s)
(blank if there is no radio position);

\item the J2000 radio interferometry declination (h,m,s)
(blank if there is no radio position).

 \end{enumerate}

   \begin{figure*}
   \centering
   \includegraphics[width=18cm]{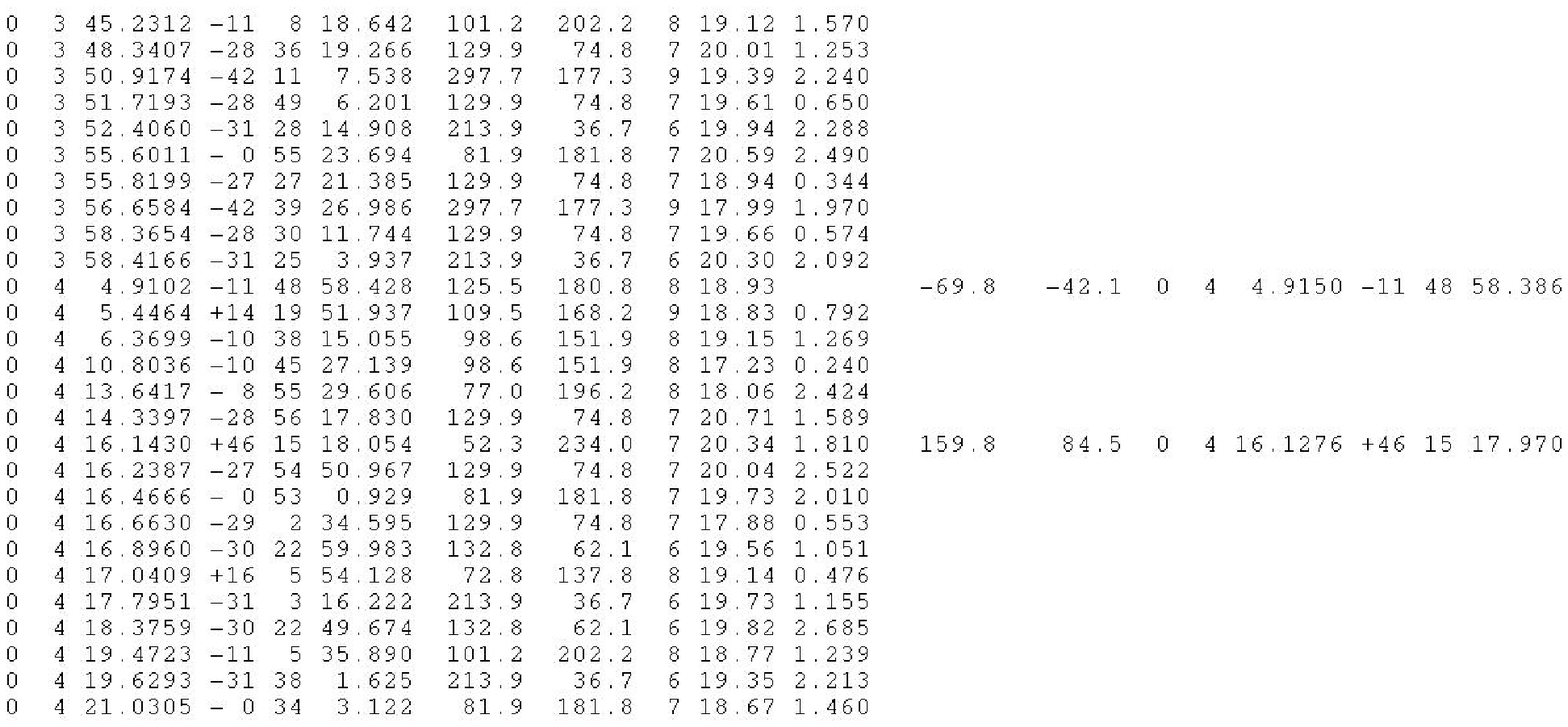}
   \caption{Example extract of the LQRF catalog. 
   Its columns contain the elements as follows.
The J2000 right ascension (h,m,s). 
The J2000 declination (d,',"). 
The external right ascension error ($\times$ cos$\delta$) (mas).
The external declination error (mas).
The number of radio interferometry neighbor positions used to evaluate 
the external errors.
The R magnitude, taken from the LQAC.
The redshift, taken from the LQAC (blank if not defined)
The right ascension offsets to the radio interferometry position (blank
if there is no radio position) ($\times$ cos$\delta$) (mas).
The declination offsets to the radio interferometry position (blank
if there is no radio position) (mas).
The J2000 radio interferometry right ascension (h,m,s)
(blank if there is no radio position).
The J2000 radio interferometry declination (h,m,s)
(blank if there is no radio position).
}
              \label{Fig25}%
    \end{figure*}

\section{Summary and perspectives}

The LQRF is build to derive an optical representation of the ICRF, retaining the 
ICRS concept of being defined by extragalactic objects. Starting from the LQAC 
entries, quasars were indentified in the USNO B1.0, GSC2.3, and SDSS DR5 catalogs.
The positions there assigned were made homogeneous with respect to the HCRF by 
applying local 
corrections obtained in small neighborhods around the quasars, relative to the 
UCAC and 2MASS catalogs. The global orientation was then adjusted to the ICRF and the
zonal departures  were removed. The ICRF was represented by the ERF containing 
accurate long base interferometry radio positions, collected from the
ICRF-Ext2, the VCS6, and the VLACalib, and selected to precision better than
10 mas.

The final LQRF J2000 equatorial cordinates were derived by weighted averages of the
input catalog positions, once locally, zonally and globally corrected as 
indicated. The errors assigned to the LQRF coordinates reflect the departure from the ERF
.

The Large Quasar Reference Frame (LQRF) formed in this way contains 100,165 objects,
of which 2,142 have accurate radio interferometric positions.
The comparison between the LQRF positions and the ERF positions indicates 
that the overall orientation towards the ICRF, as represented by the Euclidean 
direction cosines
is A1=$+$2.1 $\pm$ 3.8 mas A2=$-$0.9 $\pm$ 3.5 mas  A3=$-$2.6 $\pm$ 3.4 mas
with zero equatorial bias to the level of 2.9 mas   .
The average offsets to the ICRF are $\Delta\alpha\it{cos}\delta$=$+$2.7 $\pm$ 
2.9 mas
and $\Delta\delta$=$+$0.3 $\pm$ 2.9 mas , with standard deviations of
$\sigma_{\alpha}$=134.3 mas     and $\sigma_{\delta}$=131.1 mas . The internal 
errors
are well described by a Poisson representation, peaking at 139 mas in right 
ascension
and at 130 mas in declination.. The LQRF is planned to be maintained and enlarged in the future,
as new versions of the LQAC and other quasar input catalogs appear, as
well as to incorporate newer versions of the UCAC2 stellar reference frame.

\begin{acknowledgements}

This work would have not existed without the effort and fundibg involved in the
projects of the USNO B1.0 (US Naval Observatory), the GSC2.3 (Space Telescope Institute
and Osservat\'orio di Torino/IBAF), the Sloan Digitized Sky Survey (Astrophysical 
Research Consortium for the Participating Institutions), the ICRF (International Earth
Rotation Service), the VLBA Calibrator Survey programs (Goddard Space Flight Center NASA), 
VLA (National Radio Astronomy Observatory), the UCAC project (U.S. Naval Observatory),
and the 2MASS-The Two Micron All Sky Survey (NASA - NSF). The authors acknowledge
their use.

A.H.A. thanks CNPq grant PQ-307126/2006-0. J.I.B.C. acknowledges grant 151392/2005-6/CNPq 
and grant E-26/100.229/2008/FAPERJ. M.A. acknowledges grant E-26/170.686/2004/FAPERJ and 
grants 306028/2005-0 and 478318/2007-3 (CNPq).

\end{acknowledgements}

\end{document}